\documentclass{article}

\usepackage[preprint]{neurips_2026}

\usepackage{iftex}
\ifPDFTeX
\usepackage[utf8]{inputenc}
\usepackage[T1]{fontenc}
\else
\usepackage{fontspec}
\usepackage{xeCJK}
\setCJKsansfont{FandolHei-Regular.otf}
\setCJKmonofont{FandolHei-Regular.otf}

\fi
\usepackage{hyperref}
\usepackage{url}
\usepackage{booktabs}
\usepackage{amsfonts}
\usepackage{amssymb}
\usepackage{nicefrac}
\usepackage{microtype}
\usepackage{graphicx}
\usepackage{multirow}
\usepackage{amsmath}
\usepackage{multicol}
\usepackage[table]{xcolor}
\usepackage{cleveref}
\usepackage[normalem]{ulem}
\usepackage{fvextra}
\usepackage{tcolorbox}
\tcbuselibrary{breakable, skins}
\usepackage{listings}

\definecolor{xiaomiblue}{HTML}{4A7BCE}
\definecolor{xiaomipaleblue}{HTML}{B8DCFE}
\definecolor{xiaomiorange}{HTML}{FFA903}
\definecolor{xiaomiteal}{HTML}{03CCA0}
\definecolor{xiaomigreen}{HTML}{50B341}
\definecolor{xiaomicoral}{HTML}{ED696D}
\definecolor{xiaomilightgray}{HTML}{AAAAA8}
\definecolor{xiaomibrightblue}{HTML}{04A3FD}
\definecolor{xiaomimedgray}{HTML}{6E6E6C}
\definecolor{xiaomiblack}{HTML}{030303}
\definecolor{xiaomired}{HTML}{ee4028}
\definecolor{salmon}{HTML}{FFA07A}
\definecolor{highlightblue}{HTML}{E1F5FE}
\definecolor{baselinegray}{HTML}{F2F2F2}

\newcommand{\cmark}{\textcolor{green!60!black}{\checkmark}}
\newcommand{\xmark}{\textcolor{red}{$\times$}}
\newcommand{\naall}{\textcolor{gray!60}{-}}

\newcommand{\second}[1]{\underline{#1}}

\title{Dasheng AudioGen: A Unified Model for Generating Coherent Audio Scenes from Text}

\author{%
  Jiahao Mei$^{1,2}$
  \quad Heinrich Dinkel$^{2}$
  \quad Yadong Niu$^{2}$
  \quad Xingwei Sun$^{2}$
  \quad Gang Li$^{2}$
  \\
  \textbf{Yifan Liao}$^{2}$ \quad \textbf{Jiahao Zhou}$^{2}$ \quad \textbf{Junbo Zhang}$^{2}$  \quad \textbf{Jian Luan}$^{2}$ \quad \textbf{Mengyue Wu}$^{1}$\\
  \vspace{0.2cm} \\
$^{1}$X-LANCE Lab, Shanghai Jiao Tong University, Shanghai, China\\
$^{2}$MiLM Plus, Xiaomi Inc., Beijing, China\\
  \texttt{mengyuewu@sjtu.edu.cn}, \quad \texttt{dinkelheinrich@xiaomi.com}, \quad \texttt{zhangjunbo1@xiaomi.com}\\
}

\begin{document}

\maketitle

\begin{abstract}

Audio generation has long been fragmented, 
with speech, music, and sound effects 
produced by domain-specific models 
that fail to jointly generate coherent audio scenes from a single description. 
The key obstacles are 
insufficient fine-grained supervision for real-world mixed audio 
and limited acoustic representations for modeling concurrent audio components.
We present Dasheng AudioGen, a unified framework for generating general 
mixed-audio scenes from text. 
Dasheng AudioGen introduces structured multi-view captions, 
which explicitly decouple complex acoustic scenes into 
complementary description views, thereby enabling 
fine-grained control over audio layers. 
Furthermore, we employ a high-dimensional unified semantic-acoustic representation 
as the shared latent space. 
It injects semantic priors that facilitate cross-modal 
training convergence, while its high-dimensional 
feature space provides sufficient capacity 
to disentangle and fuse concurrent audio components effectively.
With these designs, a simple flow-matching DiT achieves high-quality end-to-end audio scene generation.
We also establish a comprehensive evaluation pipeline for audio scene generation. 
Experiments demonstrate that Dasheng AudioGen achieves 
performance approaching real-world recordings in mixed-audio categories, 
while remaining competitive with specialized models in single-type 
generation tasks. Demos are available at 
\href{https://nieeim.github.io/Dasheng-AudioGen-Web/}{\textcolor{cyan}{\textit{https://nieeim.github.io/Dasheng-AudioGen-Web/}}}.

\end{abstract}

\section{Introduction}

Current audio generation research is largely divided into 
separate domains, with independent architectures, methods, and datasets 
for speech, music, and sound effects. 
Text-to-speech (TTS) models synthesize clean speech without modeling the 
acoustic environment~\cite{hu2026qwen3tts,ren2019fastspeech}; 
text-to-music (TTM) models generate instrumental music~\cite{copet2023musicgen,agostinelli2023musiclm}; 
and text-to-audio (TTA) models generate sound effects~\cite{liu2023audioldm,liu2023audioldm2,hung2024tangoflux} but not intelligible speech. 
However, real-world audio rarely occurs as a single domain. 
For example, a news broadcast combines speech, background music, transition effects, 
and ambient sounds into one 
coherent \emph{audio scene}\footnote{In this paper, we use \emph{audio scene} to denote 
a coherent mixed-audio clip that may contain intelligible speech, music, and sound 
effects.}. 
This requires general audio generation systems to jointly model temporal relations, energy balance, masking effects, environmental consistency, and overall realism among different audio components within the same scene.
To the best of our knowledge, Dasheng AudioGen is the first non-autoregressive unified text-to-audio model explicitly designed and evaluated for coherent mixed-audio scene generation with intelligible speech, music, and sound effects in a single audio clip.

\begin{table}[t]
\centering
\caption{Generation capabilities of representative audio generation models.}
\label{tab:generation_capabilities}
\small
\begin{tabular}{lcccc}
\toprule
Model & Sound Effect & Music & Intelligible Speech & Audio Scene \\
\midrule
MusicGen~\cite{copet2023musicgen} & \xmark & \cmark & \xmark & \xmark \\
TangoFlux~\cite{hung2024tangoflux} & \cmark & \cmark & \xmark & \xmark \\
Qwen3-TTS~\cite{hu2026qwen3tts} & \xmark & \xmark & \cmark & \xmark \\
UniFlow-Audio~\cite{xu2025uniflow} & \cmark & \cmark & \cmark & \xmark \\
Dasheng AudioGen & \cmark & \cmark & \cmark & \cmark \\
\bottomrule
\end{tabular}
\end{table}

Unified audio scene generation faces two key challenges. 
The first is data and textual supervision. 
Existing high-quality audio datasets are typically domain-specific. 
For example, TTS datasets such as LibriTTS~\cite{LibriTTS} provide accurate transcripts 
and clean speech, but contain little music, sound events, or ambient sounds, thereby lacking acoustic diversity.
In contrast, in-the-wild audio contains richer acoustic scenes, but is 
often annotated only with coarse global captions. For complex mixed audio, 
a global caption is insufficient to provide the fine-grained supervision needed 
for controllable and coherent scene generation.

The second challenge is audio representation. General audio scenes contain 
heterogeneous and overlapping sound sources, making them harder to model than 
single-domain TTS, TTM, or TTA. Traditional generation systems use low-dimensional 
VAE acoustic latents as targets, 
forcing the model to learn a difficult mapping from semantic text conditions 
to low-level acoustic spaces. Such compact latents may also lack the capacity 
to represent multiple coexisting audio components and their interactions. 
Thus, unified audio generation requires not only a unified architecture, 
but also a high-capacity representation space that preserves both semantic 
structure and acoustic details.

To address these challenges, we propose Dasheng AudioGen, 
a unified text-to-audio framework for general audio scene generation. 
Our key insight is that unified generation does not require separate modules 
for different sound types; instead, it requires structured conditioning 
and a unified semantic-acoustic latent space. Specifically, 
we introduce structured multi-view captions, which decompose a complex audio 
scene into complementary textual views, including global scene description, 
speaker style, speech transcript, sound events, music description, and acoustic 
environment. Compared with a single caption, this format provides finer-grained 
supervision and control. We further adopt a unified semantic-acoustic representation 
based on DashengTokenizer, which reduces the difficulty of text-to-audio mapping 
and provides sufficient capacity for jointly modeling speech, music, and sound effects. 
With these designs, a simple flow-matching DiT achieves high-quality end-to-end audio 
scene generation.
Table~\ref{tab:generation_capabilities} compares the generation capabilities of 
representative models; Dasheng AudioGen is the only model that supports audio, 
music, intelligible speech, and coherent audio scene generation.

To systematically evaluate unified audio generation, 
we build a comprehensive evaluation pipeline covering both single-type and complex 
mixed category generation. In addition to standard benchmarks such as 
AudioCaps~\cite{kim2019audiocaps}, MusicCaps~\cite{agostinelli2023musiclm}, and 
LibriTTS~\cite{LibriTTS}, we evaluate single-type and mixed-type audio scene 
generation on MECAT~\cite{nuu2025mecat} and construct a strong expert-pipeline baseline. 
We further conduct human evaluation and 
introduce PAFI (Physical Acoustic Fidelity Index), an LLM-as-a-judge metric, 
to assess perceptual quality, text relevance, and scene realism.

Our contributions are summarized as follows:
\begin{enumerate}
    \item \textbf{End-to-end audio scene generation.}
    We propose Dasheng AudioGen, a unified 
    text-to-audio framework that 
    jointly generates intelligible speech, music, 
    sound effects, and environmental acoustics 
    within one audio scene, 
    without domain-specific modeling.
    \item \textbf{Structured multi-view captions.} We introduce a layered caption design that provides fine-grained supervision and disentangled control over different audio components, while remaining naturally compatible with agentic systems.
    \item \textbf{Generation with semantic-acoustic representation.} 
    Instead of relying on low-dimensional acoustic VAEs, 
    we introduce a unified semantic-acoustic representation based on 
    DashengTokenizer~\cite{dinkel2026dashengtokenizer} as the shared latent space for flow matching.
    It provides semantic priors for efficient training and sufficient capacity to disentangle and fuse concurrent audio components.
    \item \textbf{Evaluation pipeline for audio scene generation.} 
    We establish a comprehensive evaluation pipeline for audio scene generation. 
    Experiments show that Dasheng AudioGen substantially outperforms Expert-Pipeline in 
    mixed-audio scenes while remaining competitive with specialized models.
\end{enumerate}


\section{Related Work}

\paragraph{Text-to-Speech.}

TTS has progressed from concatenative and statistical parametric synthesis to neural models like Tacotron~\cite{shen2018tacotron}, FastSpeech~\cite{ren2019fastspeech}, VITS~\cite{kim2021vits}, and more recent systems such as Qwen3-TTS~\cite{hu2026qwen3tts}.
Recent systems achieve high naturalness on clean speech, but typically ignore acoustic context---generated speech sounds like a studio recording regardless of the described environment.

\paragraph{Text-to-Music.}
Text-to-music generation follows two dominant paradigms.
Autoregressive language models operate over discrete audio tokens: MusicGen~\cite{copet2023musicgen} uses a single-stage transformer to predict EnCodec tokens from text, while MusicLM~\cite{agostinelli2023musiclm} generates music via hierarchical sequence-to-sequence modeling.
Diffusion-based approaches instead operate in continuous latent spaces: AudioLDM2~\cite{liu2023audioldm2} applies latent diffusion with a joint audio-text representation, and JEN-1~\cite{li2023jen} uses flow matching for efficient generation.
These systems produce high-quality music but cannot generate speech or integrate with spoken content.

\paragraph{Text-to-Audio.}
Text-to-audio generation for general sound effects splits into diffusion and flow-matching approaches.
AudioLDM~\cite{liu2023audioldm} and Make-An-Audio~\cite{huang2023makeanaudio} use latent diffusion to generate sound effects from text, operating on mel-spectrogram or VAE latents.
More recently, flow-matching methods have emerged as efficient alternatives: TangoFlux~\cite{hung2024tangoflux} uses conditional flow matching for fast, high-quality TTA without the iterative sampling overhead of diffusion.
Both paradigms focus exclusively on non-speech, non-music sounds.

\paragraph{Unified Audio Generation.}
AudioX~\cite{tian2025audiox}, UniAudio~\cite{chung2024uniaudio}, 
and UniFlow-Audio~\cite{xu2025uniflow} unify multiple 
audio generation tasks with task-specific 
conditioning modules, 
such as phoneme encoders for speech and MIDI encoders for music. 
However, these designs introduce architectural 
complexity and scale poorly to new audio types. 
They also only support multiple tasks separately. 
BagPiper~\cite{tian2026bagpiper} adopts an autoregressive framework for joint audio understanding and generation, and demonstrates the ability to generate mixed audio compositions. However, it remains closed-source and relies on very long unstructured captions, approximately 500 words for a 10-second clip, which makes fine-grained and disentangled control difficult. Moreover, BagPiper does not provide a dedicated evaluation protocol for mixed-audio generation, making direct comparison on coherent mixed-audio scenes difficult.

To the best of our knowledge, Dasheng AudioGen is the first non-autoregressive unified text-to-audio model explicitly designed and evaluated for coherent mixed-audio scene generation with intelligible speech, music, and sound effects in a single audio clip.

\section{Method}

\begin{figure}[ht]
    \centering
    \includegraphics[width=0.95\linewidth]{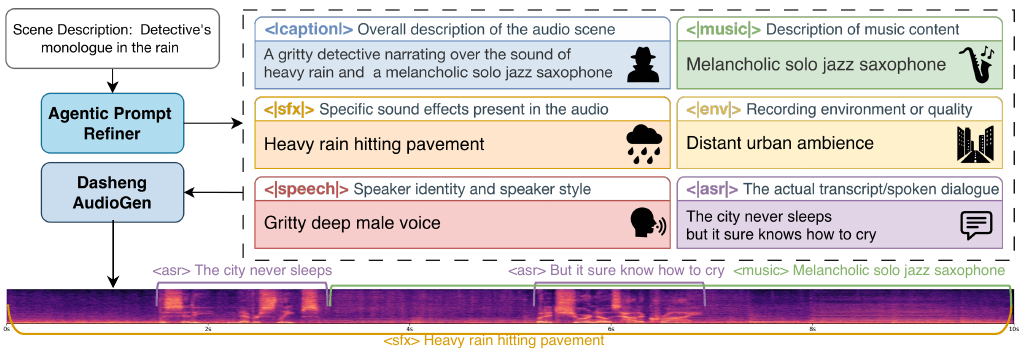}
  \caption{Structured multi-view audio scene captioning and agentic inference pipeline. Special tokens such as \texttt{<|music|>} describe different components of the target audio scene. An agentic prompt refiner automatically converts a simple scene description into a structured
  caption for fine-grained control.}    
  \label{fig:caption_view}
\end{figure}

Given a text description $y$, our goal is to generate an audio scene $x$, 
where $x$ may simultaneously contain speech, music, sound effects, 
and environmental 
acoustics. We formulate this task as conditional generation 
$p_\theta(x \mid y)$. 
\Cref{fig:caption_view} illustrates the design of structured multi-view captions and the 
agentic inference pipeline.

\subsection{Structured Multi-View Audio Scene Captioning}
\label{ssec:multi_view_caption}

Prior audio generation systems usually rely on a single coarse text prompt or label, forcing the model to infer multiple audio layers from an entangled global description. For unified audio generation, such coarse conditioning limits fine-grained control and joint modeling of different audio components.

To address this, we introduce structured multi-view captions, which decompose an audio scene into six complementary views, as shown in~\Cref{fig:caption_view}. Each view is associated with a dedicated special token, such as \texttt{<|caption|>} and \texttt{<|speech|>}. Every sample contains the \texttt{<|caption|>} field for the global scene description, while the remaining fields are used only when applicable. For example, pure speech samples do not contain \texttt{<|music|>} or \texttt{<|sfx|>}.

This structured condition provides factorized supervision 
for complex audio scenes. Instead of using separate text encoders or 
task-specific modules for different sound types, we expose different semantic 
views to a unified text encoder via explicit special tokens. 
Compared with a single global caption, multi-view captions reduce 
semantic entanglement among control factors and enable fine-grained control.

At inference, this format is also naturally compatible with large language models: given a short scene description, an LLM can populate each field separately, enabling an agentic inference interface.

\subsection{View-Aware Conditioning}

Dasheng AudioGen uses a single T5~\cite{chung2024scaling} text encoder to 
condition the flow-matching DiT. 
Given a structured multi-view caption, 
we represent it as a sequence of view segments $y = [s_1, y_1, s_2, y_2, \ldots, s_K, y_K]$, 
where $s_k$ is the special token for the $k$-th view, such as \texttt{<|asr|>},
and $y_k$ is its textual content. 
These special tokens define explicit semantic boundaries and expose view identity 
to the text encoder. The full sequence is encoded by T5 as $C\in \mathbb{R}^{L \times d_c}$, where $L$ is the number of text tokens and $d_c$ is the text embedding dimension.

The DiT generates DashengTokenizer latent sequences. Let $H_l \in \mathbb{R}^{T \times d}$ denote the audio hidden states at the $l$-th DiT block, where $T$ is the latent length and $d$ is the hidden dimension. Each block uses self-attention to model temporal dependencies among audio latents and cross-attention to inject multi-view text conditions:
\begin{equation}
\mathrm{CrossAttn}(H_l, C)
=
\mathrm{softmax}
\left(
\frac{Q(H_l)K(C)^\top}{\sqrt{d}}
\right)V(C).
\end{equation}
This allows each audio latent token to softly select relevant information from 
different caption views according to the current generation state. 
Dasheng AudioGen thus achieves view-aware conditioning with only special tokens and cross-attention, without view-specific encoders or task-specific modules.

\subsection{Semantic-Acoustic Latent Space}

Many prior models~\cite{liu2023audioldm2,xu2025uniflow} use low-dimensional
acoustic VAE latents as the generation space. However, general audio scenes
contain concurrent heterogeneous components and require joint modeling of their
interactions. This makes it difficult to map semantic text conditions to purely
acoustic latents, while the low-dimensional VAE bottleneck may discard details
needed for overlapping components.

To introduce semantic priors into the generation space, 
we use the unified semantic-acoustic representation from 
DashengTokenizer~\cite{dinkel2026dashengtokenizer}.
Given an audio waveform $x$, the 
DashengTokenizer encoder produces a continuous 
latent representation $z = E_{\mathrm{DS}}(x) \in \mathbb{R}^{T \times 1280}$ with a 
frame rate of 25 Hz. 
Unlike low-dimensional acoustic VAE latents, DashengTokenizer representations contain both semantic information and acoustic detail. The semantic prior shortens the cross-modal mapping from text to audio representations, while the high-dimensional space provides sufficient capacity to model overlapping audio components and their interactions.

\subsection{Flow Matching Objective}

\begin{figure}[t]
    \centering
    \includegraphics[width=0.9\linewidth]{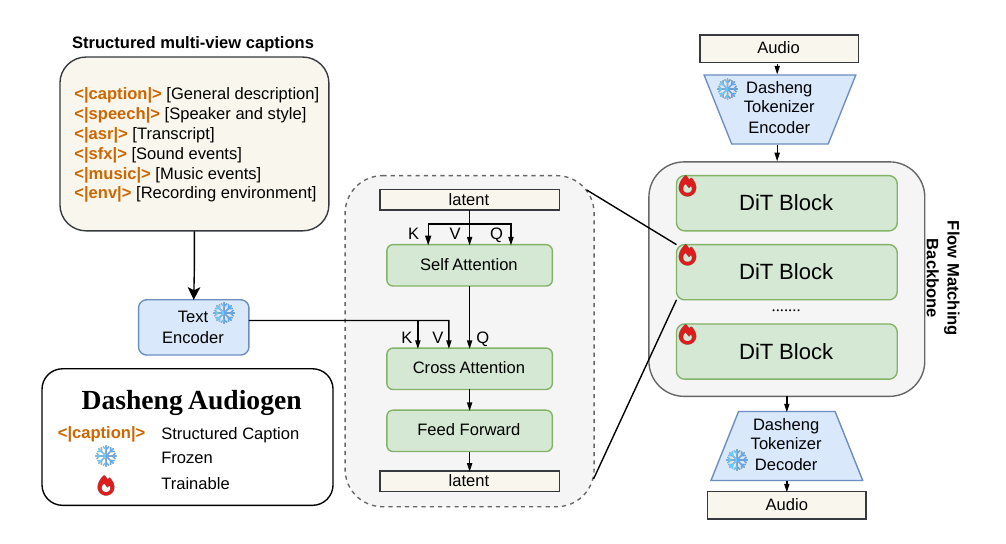}
    \caption{Overview of Dasheng AudioGen. A structured multi-view caption is encoded by T5 and used to condition a DiT that generates DashengTokenizer latents via flow matching. The DashengTokenizer decoder converts latents to waveforms.}
    \label{fig:audiogen_framework}
\end{figure}

As shown in~\Cref{fig:audiogen_framework}, we perform generation in the DashengTokenizer latent space using standard flow matching~\cite{lipman2023flow}. Let $z_1 = E_{\mathrm{DS}}(x)$ be the real audio latent and $z_0 \sim \mathcal{N}(0, I)$ be Gaussian noise. For $t \sim \mathcal{U}(0,1)$, we construct $z_t = (1-t)z_0 + t z_1$. The DiT learns a conditional vector field $v_\theta(z_t, t, C)$, where $C$ is the text condition encoded from the structured caption. The training objective is
\begin{equation}
\mathcal{L}_{\mathrm{FM}}
=
\mathbb{E}_{z_0,z_1,t,y}
\left[
\left\|
v_\theta(z_t, t, C) - (z_1-z_0)
\right\|_2^2
\right].
\end{equation}

At inference, we start from Gaussian noise and solve $d z_t / dt = v_\theta(z_t, t, C)$ 
to obtain the generated latent $\hat{z}$, which is decoded into a waveform by 
the DashengTokenizer decoder. We use classifier-free guidance for 
stronger conditioning and randomly drop caption fields during training 
to support inputs with different levels of detail.

Compared with previous diffusion-based unified models, our architecture is deliberately simple. UniAudio~\cite{chung2024uniaudio} and UniFlow-Audio~\cite{xu2025uniflow} rely on multiple task-specific encoders, which complicates optimization. In contrast, Dasheng AudioGen requires only structured audio captions as input.

\subsection{Implementation Details}
\label{ssec:appendix_implementation_details}

The DiT has width 1536 and 32 layers with approximately 2B trainable parameters.
The DashengTokenizer decoder has 173M parameters with 12 layers and hidden dimension 1280.
We use Flan-T5-Large~\cite{chung2024scaling} as the text encoder, with 780M parameters.
We train with AdamW, batch size 256, and learning rate $5 \times 10^{-4}$ with cosine decay to 10\% for 800k steps, which takes 10 days on 8 H200 GPUs.
Every multi-view caption field except \texttt{<|caption|>} is randomly dropped with probability 0.2 during training to improve robustness.
At inference, we use 25 flow-matching steps with classifier-free guidance scale 5.0.

\section{Experiments}

\subsection{Experimental Setup}

Our training dataset is ACAVCaps~\cite{acavcaps}, 
a large-scale audio captioning dataset derived 
from ACAV100M~\cite{lee2021acav100m}.
We train on a private superset with 77k hours covering speech, music, and sound effects.
ACAVCaps uses a multi-expert annotation pipeline that analyzes each audio clip from six
domain-specific perspectives, which we convert into our structured multi-view caption format.
Detailed caption construction method and examples are provided in Appendix~\ref{ssec:appendix_training_cases}.

Our main evaluation benchmark is MECAT~\cite{nuu2025mecat}, which is a held-out test set of ACAVCaps that 
categorizes audio into 
single-type (S00 = speech, 0M0 = music, 00A = sound effects) and mixed categories (0MA, S0A, SM0, SMA).
We use a compact notation where S, M, and A denote the presence of speech, music, and sound effects, respectively, and 0 denotes absence.
Unlike single-type benchmarks such as AudioCaps, MECAT contains both single and mixed-type audio samples together with rich multi-view annotations, making it especially suitable for assessing mixed audio scene generation.
We also report on AudioCaps~\cite{kim2019audiocaps}, MusicCaps~\cite{agostinelli2023musiclm}, and LibriTTS~\cite{LibriTTS} for comparability with prior work.
Across these benchmarks, we report audio distribution 
metrics~(FAD, FD, and KL), text similarity 
metrics~(CLAP~\cite{elizalde2022clap} and GLAP~\cite{dinkel2025glap}), 
and speech-related metrics~(WER and UTMOSv2~\cite{baba2024utmosv2}).
Details of the evaluation dataset statistics and metrics are provided in Appendix~\ref{ssec:appendix_objective_sample_stats} and Appendix~\ref{ssec:appendix_objective_metrics}.


For single-type audio evaluation, we compare Dasheng AudioGen against 
representative specialized models, including AudioLDM2~\cite{liu2023audioldm2}, 
TangoFlux~\cite{hung2024tangoflux}, MusicGen~\cite{copet2023musicgen}, and
Qwen3-TTS~\cite{hu2026qwen3tts}. 
We also include unified audio generation models 
that integrate multiple audio generation tasks, such as 
AudioX~\cite{tian2025audiox} and UniFlow-Audio~\cite{xu2025uniflow}. 
For mixed-audio evaluation, we further construct a 
strong Expert-Pipeline baseline. 
This baseline uses different 
expert models~(Qwen3-TTS, MusicGen, and TangoFlux) to 
generate the speech, music, and
sound-effect components of a mixed audio scene separately, 
and then mixes them into the final audio output.
Because these models prefer different input formats, 
we construct model-specific evaluation captions; 
details are provided in Appendix~\ref{ssec:appendix_objective_prompts}.

In addition, we conduct ablation studies to quantify the contribution of structured multi-view captions and the unified semantic-acoustic representation.
To complement objective metrics, we also 
conduct human evaluation and LLM evaluation 
to assess perceptual quality.

\subsection{Standard Generation Benchmarks}
\label{ssec:standard_benchmarks}

\begin{table}[t]
\centering
\caption{Standard benchmark results on AudioCaps, MusicCaps, and LibriTTS. Best in \textbf{bold} and second-best \underline{underlined}.}
\label{tab:standard_benchmarks}
\small
\resizebox{\linewidth}{!}{
\begin{tabular}{lcccccccccc}
\toprule
& \multicolumn{4}{c}{AudioCaps} & \multicolumn{4}{c}{MusicCaps} & \multicolumn{2}{c}{LibriTTS} \\
\cmidrule(lr){2-5} \cmidrule(lr){6-9} \cmidrule(lr){10-11}
Method & $\mathrm{FAD}_{\mathrm{VGG}}$ $\downarrow$ & $\mathrm{FD}_{\mathrm{PANNS}}$ $\downarrow$ & KL $\downarrow$ & CLAP $\uparrow$ & $\mathrm{FAD}_{\mathrm{VGG}}$ $\downarrow$ & $\mathrm{FD}_{\mathrm{PANNS}}$ $\downarrow$ & KL $\downarrow$ & CLAP $\uparrow$ & WER(\%) $\downarrow$ & UTMOSv2 $\uparrow$ \\
\midrule
GT & 0 & 0 & 0 & 53.0 & 0 & 0 & 0 & 37.0 & 2.82 & 3.14 \\
AudioLDM2~\cite{liu2023audioldm2} & \second{2.29} & 21.84 & \second{1.41} & \second{47.6} & 3.13 & \naall & \second{1.20} & 30.1 & \naall & \naall \\
TangoFlux~\cite{hung2024tangoflux} & \textbf{2.26} & \second{19.13} & \textbf{1.19} & \textbf{58.0} & \naall & \naall & \naall & \naall & \naall & \naall \\
AudioX~\cite{tian2025audiox} & 2.45 & 24.69 & \naall & 44.0 & \second{1.42} & \second{18.53} & \textbf{1.12} & \textbf{38.6} & \naall & \naall \\
UniFlow-Audio~\cite{xu2025uniflow} & 5.74 & \textbf{17.18} & 1.43 & \second{47.6} & 4.05 & 27.12 & 1.87 & 24.1 & \naall & \naall \\
MusicGen~\cite{copet2023musicgen} & \naall & \naall & \naall & \naall & 3.80 & \naall & 1.31 & 28.0 & \naall & \naall \\
Qwen3-TTS~\cite{hu2026qwen3tts} & \naall & \naall & \naall & \naall & \naall & \naall & \naall & \naall & \textbf{2.15} & \textbf{3.40} \\
Ours & 3.19 & 26.06 & 1.86 & 43.8 & \textbf{1.37} & \textbf{18.45} & 1.37 & \second{33.4} & \second{10.77} & \second{3.12} \\
\bottomrule
\end{tabular}
}
\end{table}

To validate Dasheng AudioGen's capabilities on single-type audio generation, we report results on AudioCaps, MusicCaps, and LibriTTS in~\Cref{tab:standard_benchmarks}.

\noindent{\textbf{Sound Effect Generation.}} On AudioCaps, Dasheng 
AudioGen (FAD 3.19) slightly trails task-optimized models such as 
AudioLDM2 (2.29) and TangoFlux (2.26), but performs comparably to 
the unified multi-task model AudioX (2.45) and substantially 
outperforms UniFlow-Audio (5.74). This performance gap can 
be attributed to three factors: (1) unlike our model, the other TTA baselines 
include AudioCaps in their training sets, yielding an in-domain advantage; 
(2) pure sound effects (00A) comprise only 1.34\% of our training 
data (Appendix Table~\ref{tab:training_data_stats}), leading 
to data scarcity in this category; and (3) 
our minimalist architecture does not include modality-specific inductive 
biases or targeted optimizations, such as the CLAP-oriented preference optimization used in TangoFlux.

\noindent{\textbf{Music and Speech Generation.}} Our method exhibits strong 
competitiveness in music and speech generation. On MusicCaps, Dasheng AudioGen 
achieves an FAD of 1.37, substantially outperforming AudioLDM2 (3.13) and MusicGen 
(3.80). It also slightly surpasses the unified multi-task model AudioX (1.42) and 
clearly outperforms UniFlow-Audio (4.05). 
On LibriTTS, its WER (10.77\%) is 
higher than that of the optimized Qwen3-TTS system (2.15\%). This is primarily 
because our model currently generates audio with a fixed 10-second duration, 
truncating longer texts and artificially inflating WER. On UTMOSv2, which better 
reflects overall speech perception and naturalness, 
we achieve a score of 3.12, approaching Qwen3-TTS (3.40).

In conclusion, despite being designed for complex mixed-audio scenes, Dasheng AudioGen maintains highly competitive performance across standard TTA, TTM, and TTS benchmarks, with particularly strong results in music quality and speech naturalness. This demonstrates that a unified representation and minimalist architecture do not compromise foundational single-type generation capabilities, providing a solid basis for its strong performance in mixed-audio scenarios.

\subsection{MECAT Benchmark}
\label{ssec:mecat}
\noindent To comprehensively evaluate model performance in complex mixed-audio scenes, we conduct experiments on the MECAT benchmark.

\noindent{\textbf{Single-Type Categories.}} Results on the single-type 
categories (00A, 0M0, and S00) are reported in Appendix Table~\ref{tab:mecat_single}. 
Dasheng AudioGen achieves the best performance on all 
acoustic distribution metrics (FAD, FD, and KL) across all 
three categories, substantially outperforming all baselines. In addition, 
our model attains the best or second-best results on the text-audio similarity 
metrics CLAP and GLAP. Notably, in the pure speech category S00, although our WER 
(22.96\% vs.\ 13.14\%) and UTMOSv2 (2.92 vs.\ 3.46) are lower than those of 
the Expert-Pipeline, our acoustic distribution metrics are markedly 
stronger (e.g., FAD 1.76 vs.\ 8.46). 
We attribute the Expert-Pipeline's weaker distributional performance mainly 
to its lack of environmental awareness. 
Unlike the clean studio-style speech in LibriTTS, the MECAT speech 
categories contain rich vocal and environmental details that reflect 
realistic acoustic conditions (see Appendix~\ref{ssec:appendix_training_cases} 
for representative examples). 
This indicates that Dasheng AudioGen better models speech together with its acoustic context, whereas specialized TTS systems typically ignore environmental descriptions and generate speech detached from the physical scene.

\begin{table}[t]
\centering
\caption{Results on MECAT mixed-audio categories. 
Best values are in \textbf{bold} and second-best values are 
\underline{underlined}. 0MA=Music+Audio, S0A=Speech+Audio, SM0=Speech+Music, SMA=Speech+Music+Audio.}
\label{tab:mecat_mixed}
\scriptsize
\resizebox{\linewidth}{!}{
\begin{tabular}{llccccccc}
\toprule
Category & Method & $\mathrm{FAD}_{\mathrm{VGG}}$ $\downarrow$ & $\mathrm{FD}_{\mathrm{PANNS}}$ $\downarrow$ & KL $\downarrow$ & CLAP $\uparrow$ & GLAP $\uparrow$ & WER(\%) $\downarrow$ & UTMOSv2 $\uparrow$ \\
\midrule
\multirow{5}{*}{0MA} & TangoFlux & 9.71 & 58.44 & 2.45 & 31.6 & 5.44 & \naall & \naall \\
 & MusicGen & 13.06 & 60.98 & 2.86 & 19.3 & 2.56 & \naall & \naall \\
 & Qwen3-TTS & 23.17 & 149.24 & 5.28 & -3.0 & -11.27 & \naall & \naall \\
 & Expert-Pipeline & \second{5.55} & \second{43.79} & \second{1.83} & \textbf{35.2} & \textbf{8.10} & \naall & \naall \\
 & Ours & \textbf{3.25} & \textbf{30.58} & \textbf{1.42} & \second{31.9} & \second{8.02} & \naall & \naall \\
\midrule
\multirow{5}{*}{S0A} & TangoFlux & 10.92 & 43.94 & 2.23 & \second{36.0} & 3.99 & 99.30 & 1.84 \\
  & MusicGen & 30.38 & 89.09 & 5.60 & 0.2 & -13.67 & 100.00 & 1.65 \\
  & Qwen3-TTS & 19.62 & 62.84 & 2.42 & 15.3 & -6.51 & \textbf{13.63} & \textbf{3.48} \\
  & Expert-Pipeline & \second{7.10} & \second{32.36} & \second{1.86} & \textbf{37.9} & \second{6.50} & 49.22 & 2.22 \\
  & Ours & \textbf{1.75} & \textbf{8.56} & \textbf{0.69} & \second{36.3} & \textbf{11.08} & \second{22.98} & \second{2.60} \\
\midrule
\multirow{5}{*}{SM0} & TangoFlux & 11.57 & 40.77 & 1.11 & \textbf{33.7} & 1.85 & 99.44 & 1.40 \\
  & MusicGen & 19.83 & 57.28 & 2.38 & 13.8 & -4.99 & 99.99 & 1.65 \\
  & Qwen3-TTS & 10.20 & 63.15 & 2.16 & 18.6 & -3.78 & \textbf{15.60} & \textbf{3.49} \\
  & Expert-Pipeline & \second{9.55} & \second{24.10} & \second{0.69} & 30.9 & \second{5.36} & 24.31 & 2.26 \\
  & Ours & \textbf{1.70} & \textbf{6.69} & \textbf{0.33} & \second{32.7} & \textbf{9.80} & \second{21.96} & \second{2.72} \\
\midrule
\multirow{5}{*}{SMA} & TangoFlux & 11.40 & 50.58 & 2.05 & 29.0 & 1.18 & 99.72 & 1.63 \\
  & MusicGen & 20.18 & 69.19 & 3.41 & 10.7 & -8.04 & 100.00 & 1.69 \\
  & Qwen3-TTS & 16.53 & 83.01 & 2.45 & 10.2 & -8.05 & \textbf{14.92} & \textbf{3.22} \\
  & Expert-Pipeline & \second{6.38} & \second{30.10} & \second{1.08} & \second{37.7} & \second{7.32} & 62.14 & 2.24 \\
  & Ours & \textbf{2.17} & \textbf{17.75} & \textbf{0.63} & \textbf{38.3} & \textbf{9.52} & \second{28.98} & \second{2.46} \\
\bottomrule
\end{tabular}
}
\end{table}

\noindent{\textbf{Mixed Categories.}} Table~\ref{tab:mecat_mixed} reports results on the mixed-audio categories 0MA, S0A, SM0, and SMA. Across all mixed categories, our model substantially outperforms all baselines on distributional similarity metrics (FAD, FD, and KL), while remaining competitive on CLAP, GLAP, WER, and UTMOSv2.
For example, in the most challenging SMA setting, which contains concurrent speech, music, and sound effects, Dasheng AudioGen achieves a remarkably low FAD of 2.17 together with a WER of 28.98\%. By comparison, the Expert-Pipeline reaches an FAD of 6.38 and a WER of 62.14\%, even though the standalone Qwen3-TTS baseline attains 14.92\% WER on the same category. 
This degradation indicates that independently generated components are difficult to combine into a coherent mixed-audio scene. In particular, the expert models lack global coordination, which leads to severe acoustic masking and mutual interference between speech, music, and sound effects after mixing. 
By contrast, the unified representation helps coordinate energy 
distribution and cross-component interactions at the scene level, 
leading to more natural and coherent disentanglement and fusion of overlapping audio layers.

\subsection{Ablation Experiments}
\label{ssec:ablation}

\noindent We ablate the two core designs of Dasheng AudioGen: structured multi-view captions and the unified semantic-acoustic representation.

\begin{table}[h]
\centering
\caption{Comparison between structured and unstructured captions on the non-speech MECAT categories and LibriTTS. Best in \textbf{bold}.}
\label{tab:structured_unstructured_mecat_non_speech}
\small
\resizebox{\linewidth}{!}{
\begin{tabular}{lcccccccccccc}
\toprule
& \multicolumn{6}{c}{Structured} & \multicolumn{6}{c}{Unstructured} \\
\cmidrule(lr){2-7} \cmidrule(lr){8-13}
Category & $\mathrm{FAD}_{\mathrm{VGG}}$ $\downarrow$ & $\mathrm{FD}_{\mathrm{PANNS}}$ $\downarrow$ & KL $\downarrow$ & GLAP $\uparrow$ & WER(\%) $\downarrow$ & UTMOSv2 $\uparrow$ & $\mathrm{FAD}_{\mathrm{VGG}}$ $\downarrow$ & $\mathrm{FD}_{\mathrm{PANNS}}$ $\downarrow$ & KL $\downarrow$ & GLAP $\uparrow$ & WER(\%) $\downarrow$ & UTMOSv2 $\uparrow$ \\
\midrule
00A & \textbf{4.07} & \textbf{17.86} & \textbf{1.35} & \textbf{10.59} & \naall & \naall & 4.86 & 22.73 & 1.51 & 9.82 & \naall & \naall \\
0M0 & \textbf{1.68} & \textbf{15.44} & \textbf{0.61} & 9.52 & \naall & \naall & 2.34 & 15.78 & 0.66 & \textbf{9.60} & \naall & \naall \\
0MA & \textbf{3.25} & \textbf{30.58} & \textbf{1.42} & \textbf{8.02} & \naall & \naall & 5.04 & 33.93 & 1.61 & 7.89 & \naall & \naall \\
LibriTTS & \naall & \naall & \naall & \naall & \textbf{10.77} & \textbf{3.12} & \naall & \naall & \naall & \naall & 52.0 & 2.70 \\
\bottomrule
\end{tabular}
}
\end{table}

\noindent{\textbf{Structured vs.\ Unstructured Captions.}} Table~\ref{tab:structured_unstructured_mecat_non_speech} compares structured and unstructured captions on MECAT and LibriTTS. In the unstructured setting, the model is trained with only the \texttt{<|caption|>} field. Because transcripts cannot be integrated fairly into plain unstructured text for MECAT speech subsets such as S00 and SMA, we use LibriTTS as the speech benchmark; prompt construction details are provided in Appendix~\ref{ssec:appendix_objective_prompts}. On the non-speech MECAT subsets 00A, 0M0, and 0MA, structured captions outperform unstructured captions on 11 of the 12 reported metrics; for example, on 0MA they reduce FAD from 5.04 to 3.25. On LibriTTS, structured captions reduce WER from 52.0\% to 10.77\% and improve UTMOSv2 from 2.70 to 3.12, confirming that explicit transcript conditioning is critical for intelligible speech generation.

\phantomsection\label{ssec:acoustic_vs_unified_embeddings}

\begin{figure*}[h]
\centering
\includegraphics[width=\textwidth]{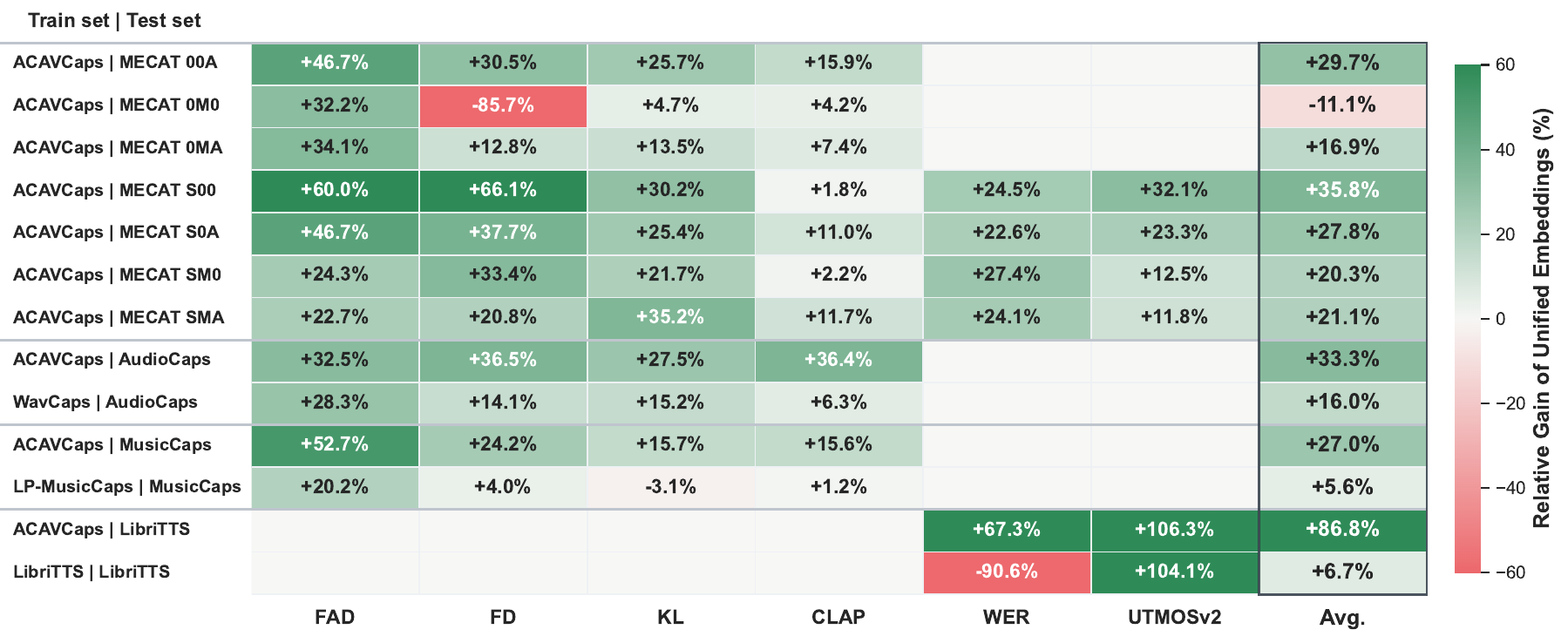}
\caption{Comparison of relative percentage gains of unified embeddings~(DashengTokenizer) 
over acoustic embeddings~(VAE). Each row is labeled as training set~$\mid$~evaluation set. }
\label{fig:acoustic_vs_unified_joint_heatmap}
\end{figure*}

\noindent{\textbf{Acoustic vs.\ Unified Embeddings.}} To validate the benefit 
of unified semantic-acoustic embedding, we train flow-matching DiT models on both 
the large-scale mixed-audio dataset ACAVCaps~\cite{acavcaps} and single-type datasets 
(WavCaps~\cite{mei2024wavcaps}, LP-MusicCaps~\cite{doh2023lp}, LibriTTS~\cite{LibriTTS}), 
using either a VAE-based acoustic representation ($d{=}128$) or DashengTokenizer's unified representation ($d{=}1280$). Figure~\ref{fig:acoustic_vs_unified_joint_heatmap} reports the percentage gain of the unified over the acoustic representation across all metrics. Detailed model configurations and absolute scores are provided in Appendix~\ref{sec:appendix_acoustic_vs_unified}.

\noindent When trained on ACAVCaps, the unified representation exhibits 
stable and substantial advantages. Across MECAT subsets, it outperforms the 
acoustic representation on nearly all metrics, with an average gain of 
approximately 20\%. Substantial improvements are also observed on the single-type evaluation sets: 
AudioCaps (+33.3\%), MusicCaps (+27.0\%), and LibriTTS (+86.8\%). This advantage cannot be explained by stronger 
acoustic reconstruction, as the unified representation does not surpass the VAE in reconstruction quality~\cite{dinkel2026dashengtokenizer}. Instead, we attribute 
it to the semantic priors embedded in the unified space, which shorten the cross-modal mapping from text to audio and facilitate more stable generation alignment in 
mixed-audio scenes. A localized exception is the FD metric on MECAT 0M0 subset, which does not alter the overall dominance of the unified representation.

\noindent By contrast, when trained on single-type datasets, the unified representation still performs better on AudioCaps (+16.0\%) and MusicCaps (+5.6\%). However, on LibriTTS, a trade-off between speech intelligibility and quality emerges: the unified representation degrades WER ($-$90.6\%) while substantially improving UTMOSv2 (+104.1\%). This contrasts with ACAVCaps training, where both WER (+67.3\%) and UTMOSv2 (+106.3\%) improve consistently. The speech quality (UTMOSv2) gain likely arises from the unified representation's larger capacity, whereas the WER behavior is less straightforward. 

Unlike TTA and TTM, which focus on global acoustic scene rendering, TTS demands strict local temporal alignment between transcripts and audio.
When trained on the clean speech dataset LibriTTS, the unified representation's semantic priors provide little additional advantage, and its higher dimensionality instead increases the difficulty of learning fine-grained pronunciation alignments. Conversely, training on mixed-audio data such as ACAVCaps requires extracting speech-relevant components from complex mixed representations and aligning them with input transcripts, significantly increasing alignment difficulty. The acoustic representation is especially vulnerable to this shift: its WER on the LibriTTS test set surges from 6.4\% to 32.9\% when training moves from clean LibriTTS to mixed ACAVCaps. The unified representation's WER on LibriTTS, however, drops from 12.2\% to 10.77\% under the same shift, suggesting that its semantic priors effectively disentangle audio layers and alleviate the alignment burden.


\begin{figure}[t]
    \centering
    \includegraphics[width=\linewidth]{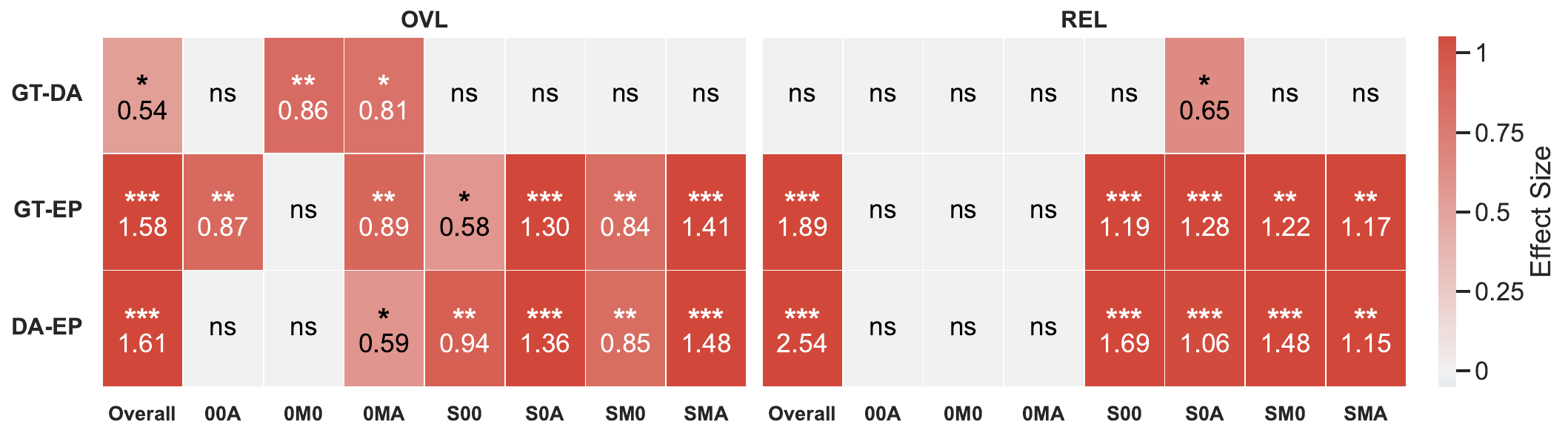}
    \caption{Pairwise human evaluation on MECAT. 
    Cells report significance and Cohen's effect size ($d_z$) for OVL and REL. 
    GT = Ground Truth, DA = Dasheng AudioGen, EP = Expert-Pipeline. 
    Significance levels, ns: $p>0.05$, *: $p\leq0.05$, **: $p\leq0.01$, ***: $p\leq0.001$.}
    \label{fig:mecat_human_eval_pairwise}
\end{figure}

\subsection{Human Evaluation and LLM Evaluation}

To complement objective evaluation, we further assess perceptual quality 
on MECAT with human evaluation and an LLM-based physical-acoustics metric~(PAFI). 
Details of the annotation protocol and supplementary statistical tests are provided in Appendix~\ref{sec:appendix_subjective_evaluation} and Appendix~\ref{sec:appendix_llm_evaluation}.

\noindent{\textbf{Human Evaluation.}} We evaluate two dimensions: 
overall quality (OVL), which measures perceived realism and the harmony 
of mixed-audio scenes, and text relevance (REL), 
as detailed in Appendix~\ref{sec:appendix_subjective_evaluation}.
We recruit 20 audio-domain professionals to rate a subset covering 
approximately 6\% of the MECAT English subset. Following a within-subject design, 
we aggregate scores at the subject level and compare systems with 
paired Wilcoxon signed-rank tests and Holm correction. 
Figure~\ref{fig:mecat_human_eval_pairwise} shows significance results and 
Appendix Figure~\ref{fig:mecat_human_eval_bar} provides mean scores.

For OVL, the gap between ground truth and Expert-Pipeline is very large on the overall set ($d_z=1.58$, $p<0.001$) and remains significant in 6 of 7 subcategories, whereas the gap between ground truth and Dasheng AudioGen is much smaller ($d_z=0.54$, $p<0.05$) and significant only in 0M0 and 0MA. Dasheng AudioGen also significantly outperforms Expert-Pipeline in all speech-mixed categories (S00, S0A, SM0, SMA) with $p < 0.01$ and $d_z > 0.85$, while showing no significant difference from ground truth.

For REL, Dasheng AudioGen differs significantly from ground truth only in S0A, matching ground-truth text relevance in the other six subcategories and overall. In contrast, the Expert-Pipeline differs significantly from both ground truth and Dasheng AudioGen on the overall set ($d_z=1.89$ and $2.54$) and across all speech-related categories.
These findings suggest that while the 
Expert-Pipeline performs adequately on 
simple non-speech tasks, it struggles with complex 
mixed-speech scenarios, which further highlights Dasheng 
AudioGen's superior controllable generation capabilities in 
complex, multi-layered scenes.

\noindent{\textbf{Physical Acoustic Fidelity Index (PAFI).}} To 
extend evaluation beyond the human-rated subset and 
assess scene coherence at a more fundamental physical acoustic level, 
we further use the LLM-based PAFI metric leveraging Gemini-3.1-Pro~(prompt shown in Appendix~\ref{ssec:appendix_pafi_prompt}). 
Appendix Figure~\ref{fig:effect_size_agreement} shows that PAFI 
aligns well with human OVL in relative system preference, with 81.0\% sign 
agreement in paired effect sizes and Pearson correlation $r=0.822$ ($p \leq 0.001$). 
Appendix Figures~\ref{fig:mecat_pafi_bar} and~\ref{fig:mecat_pafi_heatmap} report 
the mean score and significance results of PAFI on MECAT.

On the overall set, Dasheng AudioGen significantly outperforms the Expert-Pipeline (mean score 3.57 vs.\ 3.42, $d_z=0.117$, $p<0.001$), with larger gains in complex speech-containing scenes. For example, in SMA, Dasheng AudioGen reaches a PAFI score of 3.61, statistically tied with ground truth (3.60, $p>0.05$) and substantially higher than the Expert-Pipeline (2.880, $d_z=0.45$, $p<0.001$). This indicates better preservation of physically coherent interactions among overlapping audio elements.

\section{Limitations}
\label{sec:appendix_limitations}


Dasheng AudioGen has several limitations.
First, because Dasheng AudioGen's training data consists entirely of 10-second audio clips, 
the current model is limited to 10-second generation.
Second, in the TTS setting,
the model currently supports only coarse speaker-style control from 
text and does not support voice cloning or explicit speaker-identity 
conditioning. As a result, speaker similarity 
metrics are not applicable. 
Moreover, although the generated speech is natural, its intelligibility still 
lags behind specialized TTS systems.
Third, full reproducibility is currently limited because 
training relies on a much larger private superset of ACAVCaps~\cite{acavcaps} 
rather than the public release of approximately 10K hours.

\section{Conclusion}


We presented Dasheng AudioGen, 
a unified text-to-audio model for coherent audio scene generation. 
By introducing structured multi-view captions and 
semantic-acoustic latents, 
Dasheng AudioGen achieves high-quality end-to-end complex audio scene generation
using a simple flow-matching DiT framework. 
Comprehensive evaluations show 
that our method substantially outperforms expert pipelines in 
complex mixed-audio scenes while remaining competitive on single-domain tasks.
Future work will explore variable-length generation, improved speech intelligibility, and finer-grained controllability such as audio editing and explicit temporal control.
\clearpage

\bibliographystyle{unsrt}
\bibliography{references}

\clearpage
\appendix

\setcounter{table}{0}
\setcounter{figure}{0}
\renewcommand{\thetable}{A\arabic{table}}
\renewcommand{\thefigure}{A\arabic{figure}}
\renewcommand{\theHtable}{appendix.\arabic{table}}
\renewcommand{\theHfigure}{appendix.\arabic{figure}}

\section{MECAT Single-Type Results}
\label{sec:appendix_mecat_single}

\begin{table}[h]
\centering
\caption{Results on MECAT single-type categories. 
Best values are in \textbf{bold} and second-best values are \underline{underlined}. 
S00 denotes speech only audio category, 0M0 denotes music only, and 00A denotes sound effects only.}
\label{tab:mecat_single}
\scriptsize
\resizebox{\linewidth}{!}{
\begin{tabular}{llccccccc}
\toprule
Category & Method & $\mathrm{FAD}_{\mathrm{VGG}}$ $\downarrow$ & $\mathrm{FD}_{\mathrm{PANNS}}$ $\downarrow$ & KL $\downarrow$ & CLAP $\uparrow$ & GLAP $\uparrow$ & WER(\%) $\downarrow$ & UTMOSv2 $\uparrow$ \\
\midrule
\multirow{5}{*}{00A} & TangoFlux & 15.68 & 48.81 & 2.30 & 29.9 & 4.89 & \naall & \naall \\
  & MusicGen & 43.68 & 87.41 & 5.99 & -2.8 & -5.78 & \naall & \naall \\
  & Qwen3-TTS & 51.42 & 153.55 & 5.90 & -3.5 & -8.56 & \naall & \naall \\
  & Expert-Pipeline & \second{7.35} & \second{29.38} & \second{2.17} & \textbf{38.1} & \second{9.56} & \naall & \naall \\
  & Ours & \textbf{4.07} & \textbf{17.86} & \textbf{1.35} & \second{36.5} & \textbf{10.59} & \naall & \naall \\
\midrule
\multirow{5}{*}{0M0} & TangoFlux & 5.04 & 24.97 & \second{0.82} & \textbf{32.2} & \second{6.70} & \naall & \naall \\
  & MusicGen & 4.53 & 19.75 & 0.88 & 26.0 & 5.77 & \naall & \naall \\
  & Qwen3-TTS & 21.68 & 137.10 & 4.99 & -2.7 & -8.79 & \naall & \naall \\
  & Expert-Pipeline & \second{3.93} & \second{16.77} & 0.87 & 24.1 & 6.50 & \naall & \naall \\
  & Ours & \textbf{1.68} & \textbf{15.44} & \textbf{0.61} & \second{27.0} & \textbf{9.52} & \naall & \naall \\
\midrule
\multirow{5}{*}{S00} & TangoFlux & 10.54 & 53.15 & 1.63 & 29.1 & -0.91 & 99.03 & 1.71 \\
  & MusicGen & 26.74 & 89.71 & 5.38 & 4.4 & -9.54 & 100.00 & 1.61 \\
  & Qwen3-TTS & 8.75 & 27.55 & \second{0.84} & \second{31.4} & \second{0.20} & \second{14.82} & \textbf{3.60} \\
  & Expert-Pipeline & \second{8.46} & \second{25.84} & \second{0.84} & 28.4 & -0.89 & \textbf{13.14} & \second{3.46} \\
  & Ours & \textbf{1.76} & \textbf{3.93} & \textbf{0.40} & \textbf{33.9} & \textbf{11.45} & 22.96 & 2.92 \\
\bottomrule
\end{tabular}
}
\end{table}

Table~\ref{tab:mecat_single} shows the results on MECAT 
single-type categories.
Dasheng AudioGen achieves the best results on all audio distribution metrics 
(FAD, FD, and KL), ranks first or second on the text relevance 
metrics (CLAP and GLAP), and remains competitive with specialized models 
on the speech-related metrics (WER and UTMOSv2). 
Overall, these results indicate that Dasheng AudioGen 
matches or surpasses specialized models on single-type audio 
generation tasks.

\section{Acoustic vs. Unified Embedding Ablation Results}
\label{sec:appendix_acoustic_vs_unified}


\begin{table}[h]
\centering
\caption{Acoustic~(VAE) vs.\ unified~(DashengTokenizer) embedding results under 
ACAVCaps and single-type dataset training. Better value is highlighted in \textbf{bold}.}
\label{tab:appendix_embed_trainset_results}
\scriptsize
\resizebox{\linewidth}{!}{
\begin{tabular}{ll cc cc cc cc cc cc}
\toprule
& & \multicolumn{2}{c}{$\mathrm{FAD}_{\mathrm{VGG}}$ $\downarrow$} & \multicolumn{2}{c}{$\mathrm{FD}_{\mathrm{PANNS}}$ $\downarrow$} & \multicolumn{2}{c}{KL $\downarrow$} & \multicolumn{2}{c}{CLAP $\uparrow$} & \multicolumn{2}{c}{WER(\%) $\downarrow$} & \multicolumn{2}{c}{UTMOSv2 $\uparrow$} \\
\cmidrule(lr){3-4} \cmidrule(lr){5-6} \cmidrule(lr){7-8} \cmidrule(lr){9-10} \cmidrule(lr){11-12} \cmidrule(lr){13-14}
Train set & Test set & Acou. & Uni. & Acou. & Uni. & Acou. & Uni. & Acou. & Uni. & Acou. & Uni. & Acou. & Uni. \\
\midrule
ACAVCaps & MECAT-00A & 7.63 & \textbf{4.07} & 25.69 & \textbf{17.86} & 1.82 & \textbf{1.35} & 31.5 & \textbf{36.5} & \naall & \naall & \naall & \naall \\
ACAVCaps & MECAT-0M0 & 2.48 & \textbf{1.68} & \textbf{8.32} & 15.44 & 0.64 & \textbf{0.61} & 25.9 & \textbf{27.0} & \naall & \naall & \naall & \naall \\
ACAVCaps & MECAT-0MA & 4.93 & \textbf{3.25} & 35.07 & \textbf{30.58} & 1.64 & \textbf{1.42} & 29.7 & \textbf{31.9} & \naall & \naall & \naall & \naall \\
ACAVCaps & MECAT-S00 & 4.39 & \textbf{1.76} & 11.60 & \textbf{3.93} & 0.57 & \textbf{0.40} & 33.3 & \textbf{33.9} & 30.41 & \textbf{22.96} & 2.21 & \textbf{2.92} \\
ACAVCaps & MECAT-S0A & 3.28 & \textbf{1.75} & 13.74 & \textbf{8.56} & 0.92 & \textbf{0.69} & 32.7 & \textbf{36.3} & 29.68 & \textbf{22.98} & 2.11 & \textbf{2.60} \\
ACAVCaps & MECAT-SM0 & 2.24 & \textbf{1.70} & 10.04 & \textbf{6.69} & 0.43 & \textbf{0.33} & 32.0 & \textbf{32.7} & 30.24 & \textbf{21.96} & 2.42 & \textbf{2.72} \\
ACAVCaps & MECAT-SMA & 2.81 & \textbf{2.17} & 22.42 & \textbf{17.75} & 0.98 & \textbf{0.63} & 34.3 & \textbf{38.3} & 38.17 & \textbf{28.98} & 2.20 & \textbf{2.46} \\
\midrule
ACAVCaps & AudioCaps & 4.73 & \textbf{3.19} & 41.06 & \textbf{26.06} & 2.56 & \textbf{1.86} & 32.1 & \textbf{43.8} & \naall & \naall & \naall & \naall \\
WavCaps & AudioCaps & 4.27 & \textbf{3.06} & 24.72 & \textbf{21.23} & 1.84 & \textbf{1.56} & 41.2 & \textbf{43.8} & \naall & \naall & \naall & \naall \\
\midrule
ACAVCaps & MusicCaps & 2.90 & \textbf{1.37} & 24.34 & \textbf{18.45} & 1.62 & \textbf{1.37} & 28.9 & \textbf{33.4} & \naall & \naall & \naall & \naall \\
LP-MusicCaps & MusicCaps & 4.35 & \textbf{3.47} & 23.68 & \textbf{22.74} & \textbf{1.94} & 2.00 & 25.8 & \textbf{26.1} & \naall & \naall & \naall & \naall \\
\midrule
ACAVCaps & LibriTTS & \naall & \naall & \naall & \naall & \naall & \naall & \naall & \naall & 32.90 & \textbf{10.77} & 1.52 & \textbf{3.12} \\
LibriTTS & LibriTTS & \naall & \naall & \naall & \naall & \naall & \naall & \naall & \naall & \textbf{6.40} & 12.20 & 1.48 & \textbf{3.02} \\
\bottomrule
\end{tabular}
}
\end{table}

Table~\ref{tab:appendix_embed_trainset_results} presents the detailed metric values for acoustic and unified representations across different training and evaluation sets. When trained on ACAVCaps, the acoustic and unified representations use the same DiT backbone (32 layers, hidden size 1536, $\sim$2B parameters). When trained on single-type datasets, we select different DiT configurations with approximately 750M total parameters to match the preference of each representation. The acoustic representation uses a 24-layer DiT with hidden size 1024. For the unified representation, following the finding of RAE~\cite{zheng2025diffusion} that the DiT hidden size should exceed the representation dimensionality, we use an 11-layer DiT with hidden size 1536.


\clearpage

\section{Training Details}
\label{sec:appendix_training_data}

\subsection{Training Data Distribution}
\label{ssec:appendix_training_data_distribution}


\begin{table}[h!]
\centering
\caption{Training data statistics. 
Average word counts are computed on the English subset using structured and 
unstructured captions.}
\label{tab:training_data_stats}
\small
\begin{tabular}{lcccc}
\toprule
Category & Duration (h) & Proportion (\%) & Structured Avg. Words & Unstructured Avg. Words \\
\midrule
00A & 1036.54 & 1.34 & 28.24 & 11.55 \\
0M0 & 10492.79 & 13.52 & 35.89 & 13.19 \\
0MA & 465.28 & 0.60 & 35.97 & 13.23 \\
S00 & 37094.54 & 47.79 & 52.49 & 12.64 \\
S0A & 7601.22 & 9.79 & 49.82 & 12.89 \\
SM0 & 19447.17 & 25.05 & 55.60 & 13.61 \\
SMA & 1482.54 & 1.91 & 54.57 & 14.20 \\
\midrule
Total & 77620.07 & 100.00 & 50.42 & 13.01 \\
\bottomrule
\end{tabular}
\end{table}

We train on a superset of ACAVCaps, a 77k-hour multilingual 
mixed-audio dataset with rich annotations, where all audio 
clips are 10 seconds long.
Table~\ref{tab:training_data_stats} summarizes training data distribution 
statistics and average word counts for structured and unstructured captions 
across all categories.
The training distribution is dominated by speech-containing data, while also
including a substantial amount of mixed-audio content.
Pure speech~(S00) alone accounts for 47.79\% of the training set, and the four
speech-related categories~(S00, S0A, SM0, and SMA) together account for
84.54\%.
At the same time, the mixed-audio categories~(0MA, S0A, SM0, and SMA) account
for 37.35\% of the total data, providing broad coverage of scenes in which
multiple audio components co-occur and interact.
By contrast, the proportions of pure music~(0M0) and pure sound effects~(00A)
are smaller, at 13.52\% and 1.34\%, respectively.
This distribution exposes the model to a large amount of speech and mixed-audio data, which is
consistent with its strong performance on speech-related and mixed-audio scene 
generation tasks. 

Across categories, structured captions contain roughly two to three times as
many words as unstructured captions on average, showing that the structured
format provides substantially richer descriptive information.

\begin{table}[h]
\centering
\caption{Top training languages ranked by duration, followed by all remaining languages grouped as Other.}
\label{tab:training_language_dist}
\small
\begin{tabular}{lcc}
\toprule
Language & Duration (h) & Proportion (\%) \\
\midrule
English & 15367.80 & 58.86 \\
Spanish & 2740.96 & 10.50 \\
Portuguese & 1916.24 & 7.34 \\
Russian & 1217.39 & 4.66 \\
French & 933.91 & 3.58 \\
Japanese & 874.51 & 3.35 \\
Korean & 848.15 & 3.25 \\
German & 842.29 & 3.23 \\
Other & 1369.16 & 5.24 \\
\bottomrule
\end{tabular}
\end{table}

Table~\ref{tab:training_language_dist} shows the speech language distribution
of training data.
English is the dominant language in the training data, accounting for 58.86\%,
followed by Spanish, Portuguese, and Russian.
This imbalance is likely to affect model performance across languages.

\clearpage

\subsection{Structured Caption Construction Method and Training Cases}
\label{ssec:appendix_training_cases}

ACAVCaps~\cite{acavcaps} uses a multi-expert 
annotation pipeline that analyzes each audio clip from six domain-specific perspectives:
\texttt{long} and \texttt{short} (detailed and summarized scene descriptions), 
\texttt{speech} (speaker characteristics), 
\texttt{music} (music description), 
\texttt{sound} (sound events and effects), 
and \texttt{environment} (acoustic properties such as reverberation and recording quality).
This multi-perspective annotation scheme directly motivates our structured caption design.
In our training format, the \texttt{long}/\texttt{short} annotations are mapped into \texttt{<|caption|>} to 
provide an overall scene description, 
\texttt{speech} is mapped to \texttt{<|speech|>} to describe speaker style, 
\texttt{music} is mapped to \texttt{<|music|>}, \texttt{sound} is mapped to \texttt{<|sfx|>}, 
and \texttt{environment} is mapped to \texttt{<|env|>} to encode acoustic context.
For samples that contain speech, we additionally generate \texttt{<|asr|>} transcripts using Whisper~\cite{radford2023whisper}.
This conversion ensures that the model input preserves the 
annotation granularity available in the source data while matching the structured conditioning format used at training time.

Below we show structured caption training cases from different subcategories without cherry-picking.

\newcommand{\traininline}[3]{\textcolor{#1}{\texttt{#2}}\ #3}
\newcounter{traincase}[section]
\renewcommand{\thetraincase}{\arabic{traincase}}
\newcommand{\traincasebox}[3]{%
\refstepcounter{traincase}%
\begin{tcolorbox}[
    enhanced,
    breakable,
    colback=gray!5,
    colframe=gray!70!black,
    boxrule=0.8pt,
    arc=4pt,
    outer arc=4pt,
    left=8pt, right=8pt, top=8pt, bottom=8pt]
\label{#2}
\textbf{Training Case~\thetraincase: #1}\par\smallskip
\footnotesize
#3
\end{tcolorbox}}

\traincasebox{S00 (English)}{case:s00_english}{%
\traininline{xiaomired}{<|caption|>}{Male voice analyzing volleyball match outcomes with clear diction.}\enspace
\traininline{xiaomiblue}{<|speech|>}{Sportscast-style narration: Two segments discussing team performances (Italy vs. Serbia/US/China) with valence shift from neutral to positive.}\enspace
\traininline{xiaomigreen}{<|asr|>}{only two sets left to the opponents' both lost against Serbia. Italy took a second win in a row in 2011, getting the better hand on United States and China}\enspace
\traininline{xiaomiteal}{<|env|>}{Studio-quality speech recording with intermittent low-frequency interference.}}

\traincasebox{0M0}{case:0m0_random_1}{%
\traininline{xiaomired}{<|caption|>}{Upbeat acoustic guitar instrumental with background distortion.}\enspace
\traininline{xiaomiorange}{<|music|>}{Folk-inspired guitar performance with rhythmic precision and harmonic resonance.}\enspace
\traininline{xiaomiteal}{<|env|>}{Studio recording with persistent electrical interference artifacts.}}

\traincasebox{00A}{case:00a_random_1}{%
\traininline{xiaomired}{<|caption|>}{Vehicle engine idling with persistent background interference.}\enspace
\traininline{xiaomicoral}{<|sfx|>}{Mechanical engine operation with electronic interference.}\enspace
\traininline{xiaomiteal}{<|env|>}{Indeterminate recording space with electrical interference.}}

\traincasebox{SMA (English)}{case:sma_english}{%
\traininline{xiaomired}{<|caption|>}{Neutral speech utterance precedes door sound against melancholic musical accompaniment with audio artifacts.}\enspace
\traininline{xiaomiblue}{<|speech|>}{Neutral-toned utterance: `Okay, open it'.}\enspace
\traininline{xiaomigreen}{<|asr|>}{Okay, open it.}\enspace
\traininline{xiaomicoral}{<|sfx|>}{Distinct door movement noise preceding vocal utterance.}\enspace
\traininline{xiaomiteal}{<|env|>}{Variable acoustic environment with noticeable background interference.}}

\traincasebox{SMA (Portuguese)}{case:sma_portuguese}{%
\traininline{xiaomired}{<|caption|>}{Multiple speakers discussing votes over synth-guitar blend and broadcast artifacts.}\enspace
\traininline{xiaomiblue}{<|speech|>}{Multiple Portuguese speakers discussing voting preferences ('Torcerei por Rita Cadillac... meu voto também') with intermittent overlapping dialogue.}\enspace
\traininline{xiaomigreen}{<|asr|>}{vai desgastando a pessoa. Torcerei por Rita Cadillac nessa edição. Quem não torcerá é por Rita Cadillac. Ela tem o meu voto também. Também? Vamos acompanhar o voto do...}\enspace
\traininline{xiaomicoral}{<|sfx|>}{Transmission interference noises overlapping content.}\enspace
\traininline{xiaomiorange}{<|music|>}{Synthesized instrumentation blending electronic elements and guitar characteristics.}\enspace
\traininline{xiaomiteal}{<|env|>}{Broadcast booth setting exhibiting equipment noise and signal degradation.}}

\traincasebox{SMA (German)}{case:sma_german}{%
\traininline{xiaomired}{<|caption|>}{Female voice speaking over synth-driven pop music and subtle ambient sounds.}\enspace
\traininline{xiaomiblue}{<|speech|>}{Female voice explaining a hair preparation process in German.}\enspace
\traininline{xiaomigreen}{<|asr|>}{lockern und einige Strähnen rausziehen. Dann nehme ich einfach den restlichen Teil meiner Habe und schiebe es in die Hand.}\enspace
\traininline{xiaomicoral}{<|sfx|>}{Ambiguous environmental noises suggesting manual activity.}\enspace
\traininline{xiaomiorange}{<|music|>}{Electronically processed guitar patterns with pop music structure.}\enspace
\traininline{xiaomiteal}{<|env|>}{Indoor recording with mild electrical interference and background artifacts.}}

\clearpage
\section{Objective Evaluation}
\label{sec:appendix_objective_evaluation}

\subsection{Evaluation Datasets Statistics}
\label{ssec:appendix_objective_sample_stats}

We evaluate model performance on AudioCaps, MusicCaps, LibriTTS, and MECAT. 
Since Dasheng AudioGen is designed to generate 10-second audio clips, we evaluate on the subset of LibriTTS-test-clean whose utterances are shorter than 10 seconds. 
MECAT is a multilingual benchmark, and its speech categories contain utterances in multiple languages. 
To improve evaluation stability and reduce variation from multilingual ASR performance, 
we report objective results on the English subset of MECAT for the speech-related categories. 
Table~\ref{tab:objective_eval_sample_stats} summarizes the number of evaluation samples used in each dataset. 
For filtered datasets, we report the sample counts before and after filtering.
We note that metrics on categories with relatively few samples, 
such as MECAT 0MA and SMA, may be less stable and reliable than those on larger subsets.

\begin{table}[h]
\centering
\caption{Sample statistics for the objective evaluation datasets. For filtered datasets, 
we report the sample counts as ``original $\rightarrow$ used''; 
otherwise we report the original sample count directly.}
\label{tab:objective_eval_sample_stats}
\small
\begin{tabular}{lc}
\toprule
Evaluation dataset & Samples \\
\midrule
AudioCaps & 957 \\
MusicCaps & 4950 \\
LibriTTS & 4837 $\rightarrow$ 3835 \\
MECAT 00A & 848 \\
MECAT 0M0 & 2593 \\
MECAT 0MA & 199 \\
MECAT S00 & 7839 $\rightarrow$ 3838 \\
MECAT S0A & 2439 $\rightarrow$ 1341 \\
MECAT SM0 & 5312 $\rightarrow$ 2620 \\
MECAT SMA & 643 $\rightarrow$ 283 \\
MECAT & 19873 $\rightarrow$ 11722 \\
\bottomrule
\end{tabular}
\end{table}

\subsection{Objective Metrics}
\label{ssec:appendix_objective_metrics}

We use the following objective metrics for evaluation:

\noindent{\textbf{FAD \& FD}} Following previous work~\cite{liu2023audioldm2}, we report $\mathrm{FAD}_{\mathrm{VGG}}$ and $\mathrm{FD}_{\mathrm{PANNS}}$ to measure the distributional similarity between 
generated and reference audio based on VGGish~\cite{vggish} features and PANNs CNN14~\cite{kong2020panns} features, respectively. 
We compute these metrics using the AudioLDM evaluation toolkit\footnote{\url{https://github.com/haoheliu/audioldm_eval}}.

\noindent{\textbf{CLAP \& GLAP}} We use CLAP~\cite{elizalde2022clap}\footnote{\url{https://huggingface.co/lukewys/laion_clap}, using \texttt{630k-audioset-fusion-best.pt}} and GLAP~\cite{dinkel2025glap}\footnote{\url{https://huggingface.co/mispeech/GLAP}} to measure semantic relevance between generated audio and the input prompt. Both scores are computed as cosine similarity between audio and text embeddings. On MECAT, all reference texts for these metrics use the overall caption of each sample, i.e., \texttt{<|caption|>}. Compared with CLAP, GLAP is trained with broader speech and multilingual supervision, making it more sensitive to linguistic content while remaining effective for general audio-text matching.

\noindent{\textbf{WER}} We use Word Error Rate (WER) to evaluate how accurately the generated speech matches the target transcription. To reduce transcription hallucinations on acoustically complex mixed-audio samples, 
we use the NeMo ASR model\footnote{\url{https://huggingface.co/nvidia/stt_en_conformer_transducer_xlarge}} for transcription.

\noindent{\textbf{UTMOSv2}} We use UTMOSv2~\cite{baba2024utmosv2}\footnote{\url{https://github.com/sarulab-speech/UTMOSv2}} as a reference-free metric for overall speech quality assessment. Unlike reference-based metrics, UTMOSv2 predicts a perceptual quality score directly from the generated waveform and does not require paired ground-truth speech. This is particularly useful in our setting, where speech often co-occurs with music or sound effects, making ASR-based metrics more sensitive to background interference. UTMOSv2 therefore provides a complementary view of perceived speech quality beyond transcription accuracy.

For reproducibility, Table~\ref{tab:baseline_model_versions} also lists the exact Hugging Face Hub checkpoints used for all baseline systems. We include these identifiers because performance can vary substantially across different releases, scales, and checkpoint variants of the same model family.

\begin{table}[h!]
\centering
\caption{Exact model versions used for baseline comparison.}
\label{tab:baseline_model_versions}
\small
\begin{tabular}{ll}
\toprule
Model & Hugging Face Hub name \\
\midrule
AudioLDM2 & haoheliu/audioldm2-full \\
TangoFlux & declare-lab/TangoFlux  \\
MusicGen & facebook/musicgen-large  \\
Qwen3-TTS & Qwen/Qwen3-TTS-12Hz-1.7B-VoiceDesign \\
AudioX & Zeyue7/AudioX \\
UniFlow-Audio & wsntxxn/UniFlow-Audio-large \\
\bottomrule
\end{tabular}
\end{table}

\subsection{Prompt Details for Objective Evaluation}
\label{ssec:appendix_objective_prompts}

To ensure a fair evaluation, we adopt different prompt construction 
strategies for different systems and evaluation datasets, so as to better match 
each model's input format and preference. 
Below we summarize the prompt design used for each benchmark.

\noindent{\textbf{AudioCaps and MusicCaps.}}
For this benchmark group, the prompt construction is as follows:
\begin{itemize}
\item \textbf{Dasheng AudioGen:} We prepend the original dataset prompt with the special token \texttt{<|caption|>} to keep the input format consistent with the model's training setup.
\item \textbf{Other baselines:} We directly use the original text prompt provided by the dataset.
\end{itemize}

\noindent{\textbf{LibriTTS.}}
For LibriTTS, different systems use different prompt formats:
\begin{itemize}
\item \textbf{Qwen3-TTS:} We feed the transcript directly as input and leave the instruction field empty.
\item \textbf{Dasheng AudioGen (structured):} We construct a prompt containing both a scene-level description and an \texttt{<|asr|>} field. For example, when the transcript is \textit{``I can't play with you like a little boy any more,'' he said slowly.}, the structured prompt is \textit{<|caption|> Studio-quality, high-fidelity audiobook recording <|asr|> "I can't play with you like a little boy any more," he said slowly.}
\item \textbf{Dasheng AudioGen (unstructured):} We use a plain text prompt without special fields, e.g., \textit{Studio-quality, high-fidelity audiobook recording with content "I can't play with you like a little boy any more," he said slowly.}
\end{itemize}
This design improves transcript adherence while preserving clear, high-quality speech generation.

\noindent{\textbf{MECAT.}}
Because MECAT contains more complex multimodal audio scenes, prompt construction differs substantially across systems:
\begin{itemize}
\item \textbf{Dasheng AudioGen (structured):} We use the full multi-view structured caption.
\item \textbf{Dasheng AudioGen (unstructured):} We use only the content associated with \texttt{<|caption|>}.
\item \textbf{TangoFlux and MusicGen:} We use all available information from the multi-view structured caption to avoid information loss.
\item \textbf{Qwen3-TTS:} The transcript input is taken from the content of \texttt{<|asr|>}. If this field is absent, the transcript field is left empty. The instruction field is constructed by concatenating all remaining textual content other than \texttt{<|asr|>}.
\item \textbf{Expert-Pipeline:} When \texttt{<|sfx|>} is present, its content is used as the prompt for TangoFlux to generate the sound-effects track. When \texttt{<|music|>} is present, its content is used as the prompt for MusicGen to generate the music track. When \texttt{<|speech|>} is present, Qwen3-TTS is used to generate the speech track, with \texttt{<|asr|>} serving as the transcript and \texttt{<|speech|>} as the instruction. The generated tracks are aligned at the beginning and mixed into a single final audio clip. For example, for the 0MA category, the Expert-Pipeline uses the contents of \texttt{<|music|>} and \texttt{<|sfx|>} as prompts for MusicGen and TangoFlux, respectively, and then mixes the generated outputs to form the final audio.
\end{itemize}

\clearpage
\section{Subjective Evaluation}
\label{sec:appendix_subjective_evaluation}

\subsection{Human Evaluation Setting and Results}
\label{ssec:appendix_subjective_results}
\begin{figure}[h!]
    \centering
    \includegraphics[width=\linewidth]{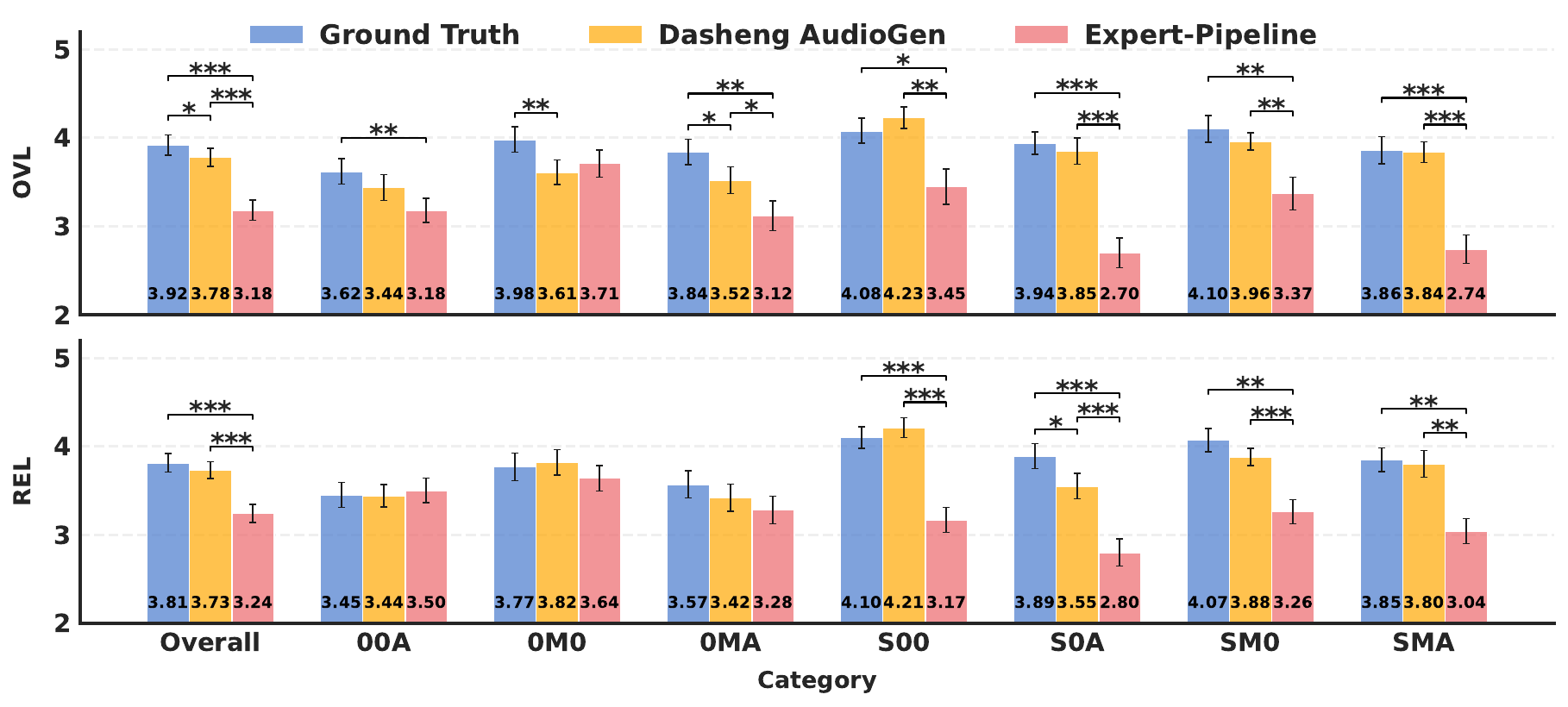}
    \caption{Human evaluation mean score and standard error~(SE) on MECAT. 
    We report subject-level mean ratings for overall quality (OVL) and relevance 
    to the prompt (REL) on the overall set and each category. 
    Error bars indicate variability across annotators after subject-level aggregation. 
    Line over two bars indicates a statistically significant difference between 
    the two systems. 
    Significance levels: *~($p \leq 0.05$), **~($p \leq 0.01$), and ***~($p \leq 0.001$).}
    \label{fig:mecat_human_eval_bar}
\end{figure}

Figure~\ref{fig:mecat_human_eval_bar} summarizes the subject-level 
mean human evaluation scores and standard error~(SE)
for different systems across all categories on both metrics (OVL and REL) in 
the MECAT benchmark. 
Main text Figure~\ref{fig:mecat_human_eval_pairwise} further presents the statistical significance and effect sizes of the human evaluation results across different categories.

We recruited 20 audio professionals
to evaluate the outputs of different systems on MECAT. Each 
participant rated 35 test trials. These 35 trials were drawn from the 7 MECAT 
subcategories,
with five trials sampled from each
subcategory. Each test trial contained three audio clips paired with the same text description: 
one generated by Dasheng AudioGen, one generated by the Expert-Pipeline, 
and the ground truth recording. 
Participants were asked to rate the audio clips in terms of OVL and REL; 
the detailed instruction and the
scoring criteria are provided in Appendix~\ref{ssec:appendix_subjective_instruction}. 
The evaluation items for each participant were 
randomly sampled from MECAT, and the final evaluation 
covered approximately 6\% of the MECAT English subset.

For statistical analysis, we conduct all comparisons at the 
subject level: for each category and each system, we first average ratings 
within each annotator and then compare systems using these subject-level means. 
We use paired Wilcoxon signed-rank tests for significance testing and apply 
Holm correction within each metric-category block to account for the three 
pairwise system comparisons.
In addition to significance, we report the paired effect size Cohen's $d_z$, 
computed from subject-wise score differences. Significance indicates whether a 
preference is statistically reliable under the current sample size, 
while effect size measures how large that preference is in practice.

\clearpage
\subsection{Human Evaluation Instruction and Rating Criteria}
\label{ssec:appendix_subjective_instruction}

Annotators were asked to rate each generated audio clip along two dimensions:
overall quality (OVL) and relevance to the input prompt (REL).
The instruction below was shown to emphasize that our target is not merely clean isolated sounds, but coherent audio scenes that sound plausibly recorded in the real world.
In particular, realism was treated as a core criterion for OVL, while REL focused on whether the generated speech, sound effects, music, environment, and overall atmosphere matched the prompt.

\newcommand{\evalinstrbox}[2]{%
\begin{tcolorbox}[
    enhanced,
    breakable,
    colback=gray!5,
    colframe=gray!70!black,
    boxrule=0.8pt,
    arc=4pt,
    outer arc=4pt,
    left=8pt, right=8pt, top=8pt, bottom=8pt]
\textbf{#1}\par\smallskip
\footnotesize
#2
\end{tcolorbox}}

\evalinstrbox{Annotation Instructions}{%
Read the text prompt and listen to the corresponding three audio samples.
Please rate each audio on overall quality (OVL) and text relevance (REL) based on the following rating criteria.
The main goal of the systems in this evaluation is to generate audio scenes with rich acoustic information that sound like they were recorded in the real world.
Please treat \textbf{REALISM} as a core criterion, and give higher scores to audio that sounds more realistic and natural.}

\evalinstrbox{Rating Criteria for OVL}{%
Please evaluate the overall quality and perceived realism of the audio.
Consider both the intrinsic quality of each sound element (speech, sound effects, music) and, when multiple sounds are present, how natural and harmonious their mix is.

\begin{enumerate}
    \item \textbf{Very poor.} The audio contains severe noise, clipping, or strong AI-generated artifacts and feels completely unrealistic. If multiple sounds are mixed, the result is extremely chaotic, heavily conflicting, and unbearable to hear.
    \item \textbf{Poor.} Individual sounds have obvious artifacts, distortion, or unnaturalness and clearly sound generated. If multiple sounds are mixed, they feel crudely stitched together, interfere with each other, or are severely imbalanced in volume, making the result uncomfortable to hear.
    \item \textbf{Average.} Individual sounds are basically clear, with occasional minor noise or AI artifacts, and only moderate realism. If multiple sounds are mixed, the parts are distinguishable and do not severely clash, but the layering is weak and the result feels ordinary.
    \item \textbf{Good.} Individual sounds are clear and natural, and the listening experience is close to a real acoustic environment. If multiple sounds are mixed, they are blended fairly well, with a clear sense of foreground and background, resulting in a comfortable and harmonious sound.
    \item \textbf{Excellent.} The audio quality is outstanding, with almost no AI artifacts and a highly realistic sound. If multiple sounds are mixed, all elements are integrated beautifully with rich layering, comparable to real-world field recordings or professional film and TV sound production, and highly immersive.
\end{enumerate}}

\evalinstrbox{Rating Criteria for REL}{%
Please evaluate how well the generated audio matches the input text, including whether it contains the requested dialogue, sound effects, music style, recording environment, and overall atmosphere.

\begin{enumerate}
    \item \textbf{Completely irrelevant.} The generated audio has nothing to do with the input text, and all required elements are wrong or missing.
    \item \textbf{Low relevance.} Most key descriptions are missing. For example, the dialogue may be completely incorrect, or the specified sound effects or music style may be absent.
    \item \textbf{Basically relevant.} The audio matches the main intent of the text and includes the major sound elements, but some specific audio elements are still missing.
    \item \textbf{Highly relevant.} Most of the text description is reproduced accurately. The speech, sound effects, music, and recording environment are basically correct, with only minor deviations in detail.
    \item \textbf{Perfect match.} All detailed requirements of the text are reproduced accurately. The elements are complete, the pronunciation is correct, and the audio fully matches the atmosphere and emotion described in the text.
\end{enumerate}}
\clearpage
\subsection{Human Evaluation Interface}
\label{ssec:appendix_subjective_interface}

\begin{figure}[h!]
    \centering
    \includegraphics[width=\linewidth]{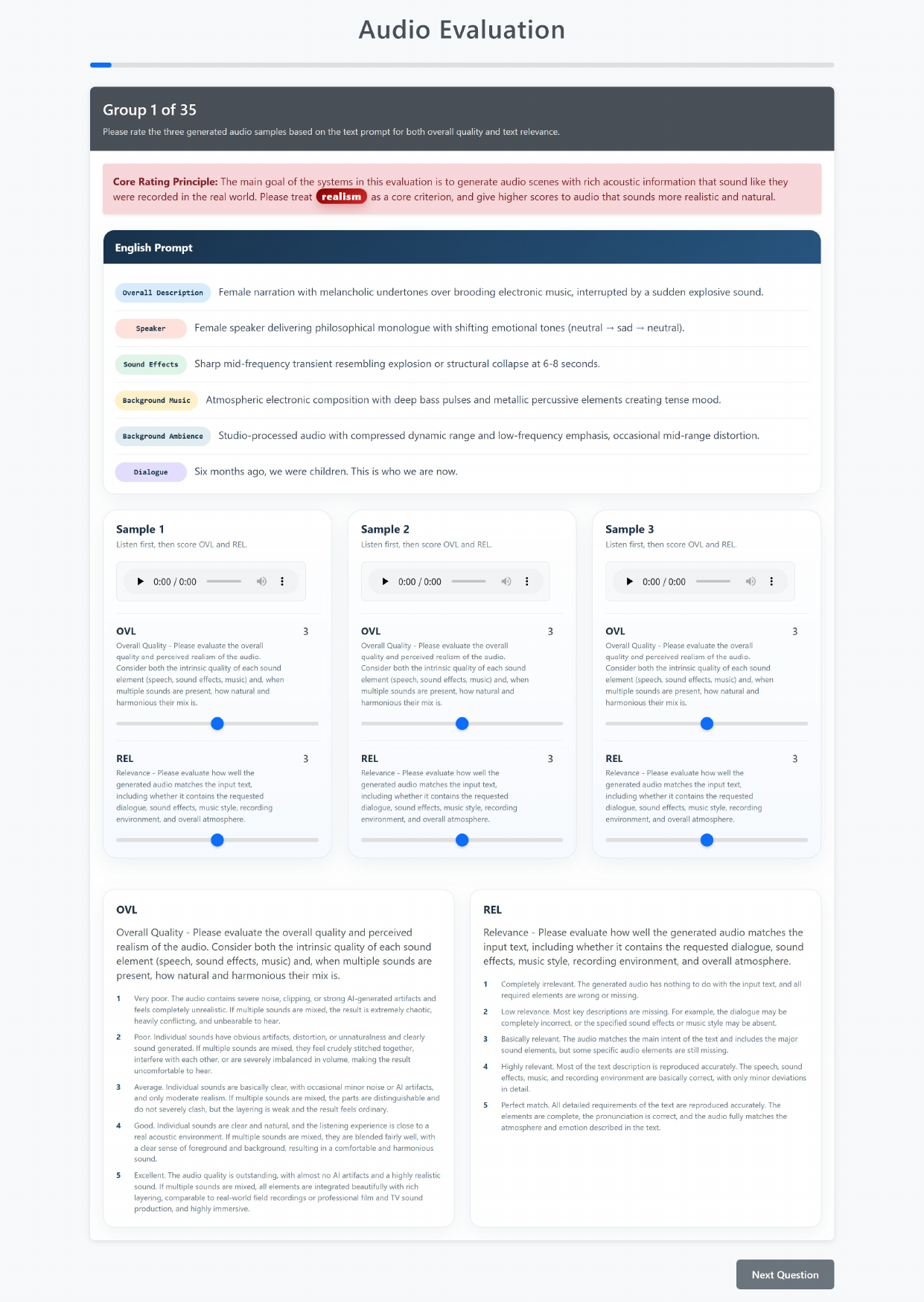}
    \caption{Human evaluation interface screenshot. 
    Annotators rated Overall Quality (OVL) and Text Relevance (REL) for each audio sample.}
    \label{fig:human_eval_ui}
\end{figure}

\clearpage

\section{LLM Evaluation}
\label{sec:appendix_llm_evaluation}

\subsection{PAFI Setting and Results}
\label{ssec:appendix_pafi_results}


\begin{figure}[h!]
    \centering
    \includegraphics[width=\linewidth]{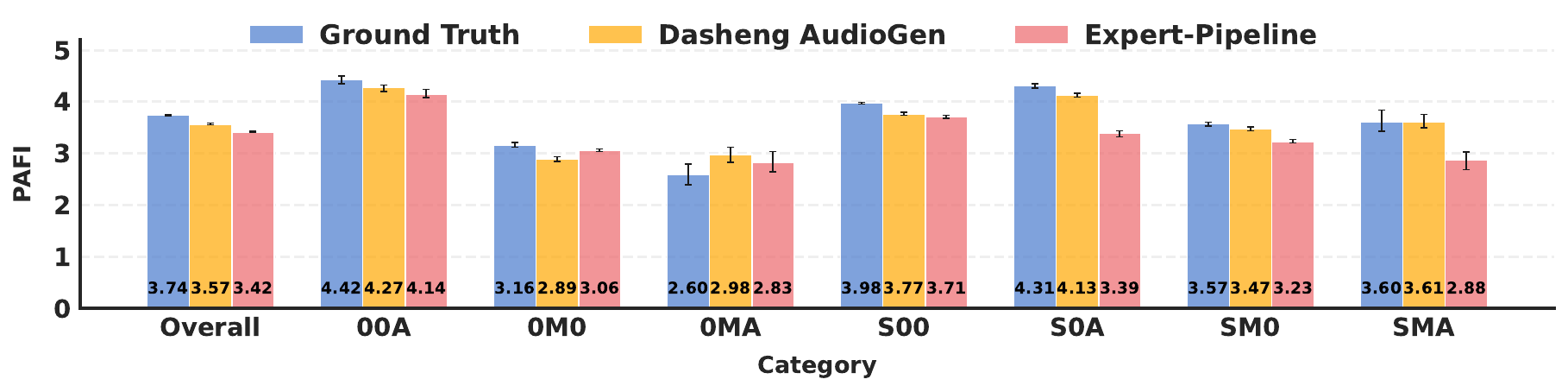}
    \caption{Mean PAFI scores and bootstrap 95\% confidence intervals on MECAT.
    Higher is better.}
    \label{fig:mecat_pafi_bar}
\end{figure}

Since human evaluation covers only a subset of MECAT, we introduce Physical Acoustic Fidelity Index~(PAFI) 
as a complementary metric for human evaluation. PAFI is an
LLM-as-a-judge metric powered by Gemini-3.1-Pro, which focuses on physical
acoustic fidelity, including spatial consistency, reverberation coherence, and
physically plausible source interaction. The prompt used for PAFI scoring is
provided in Appendix~\ref{ssec:appendix_pafi_prompt}. We report PAFI results
for Ground Truth, Dasheng AudioGen, and the Expert-Pipeline on the overall 
set and on each MECAT category.

\begin{figure}[h!]
    \centering
    \includegraphics[width=\linewidth]{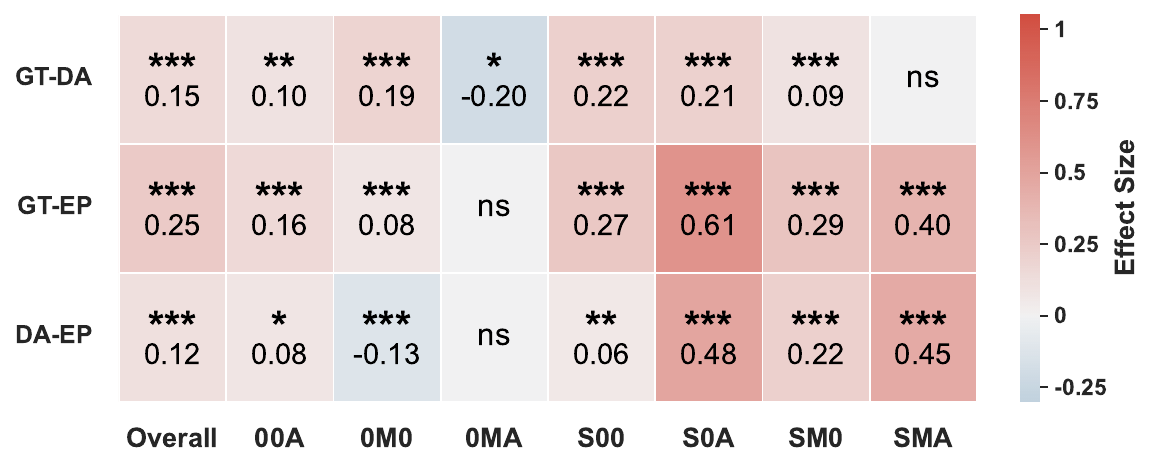}
    \caption{Pairwise significance and effect size 
    summary on PAFI. 
    Each cell compares a pair of systems within one category. 
    GT = Ground Truth, DA = Dasheng AudioGen, 
    EP = Expert-Pipeline. 
    Significance levels: ns ($p > 0.05$), * ($p \leq 0.05$), ** ($p \leq 0.01$), and *** ($p \leq 0.001$).}
    \label{fig:mecat_pafi_heatmap}
\end{figure}

Because PAFI produces a single score for each audio sample, all statistical
analyses are conducted at the sample level. 
For each category and each system
pair, we perform paired Wilcoxon
signed-rank tests, and apply Holm correction across the three pairwise
comparisons within the same category at sample-level. 
We also report the paired effect size
Cohen's $d_z$ computed from the aligned score differences.
\Cref{fig:mecat_pafi_bar} reports the mean scores together with bootstrap 95\%
confidence intervals, and \Cref{fig:mecat_pafi_heatmap} summarizes the
corresponding pairwise significance and effect sizes.

The results show a clear overall ordering. On the full evaluation set, the mean
PAFI scores are 3.74 for Ground Truth, 3.57 for Dasheng AudioGen, and 3.42 for
the Expert-Pipeline. All three overall pairwise comparisons remain significant
after Holm correction, with Dasheng AudioGen ranking between Ground Truth and
the Expert-Pipeline. The largest advantages over the Expert-Pipeline appear in
speech-containing mixed categories such as S0A and SMA. In particular, on SMA,
Dasheng AudioGen~(3.611) is nearly tied with Ground Truth~(3.604) while
remaining far above the Expert-Pipeline~(2.880). Overall, these results show
that PAFI captures meaningful system-level differences, and Appendix~\ref{ssec:appendix_pafi_human_consistency}
further shows that its system preferences are strongly aligned with human
judgment.

\subsection{PAFI-Human Evaluation Consistency}
\label{ssec:appendix_pafi_human_consistency}

\begin{figure}[h!]
    \centering
    \includegraphics[width=0.9\linewidth]{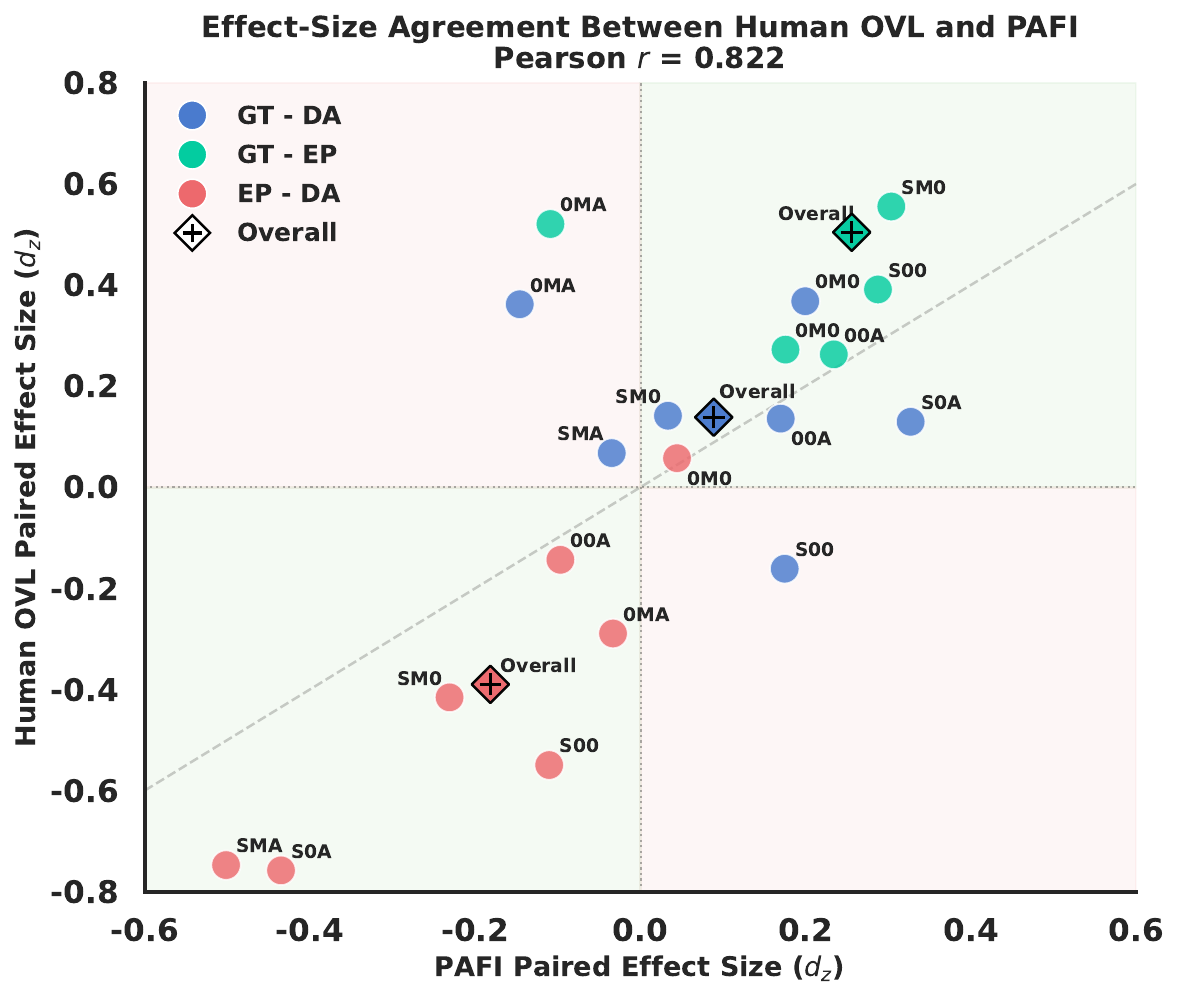}
    \caption{Agreement between human OVL and PAFI in paired 
    effect sizes. Each point corresponds to one system pair in one category. 
    The horizontal axis shows the paired effect size computed 
    from PAFI, and the vertical axis shows the corresponding 
    paired effect size computed from human OVL. 
    The diagonal dashed line indicates perfect agreement in both sign and magnitude. 
    Green-shaded quadrants indicate sign agreement, 
    and red-shaded quadrants indicate sign disagreement.
    GT-DA means Ground Truth vs.\ Dasheng AudioGen, GT-EP means Ground Truth vs.\ 
    Expert-Pipeline, and 
    EP-DA means Expert-Pipeline vs.\ Dasheng AudioGen.
    }
    \label{fig:effect_size_agreement}
\end{figure}

To evaluate whether the proposed PAFI metric aligns with human preference, we
further examine consistency at the level of \emph{relative system preference}.
Human evaluation covers only a subset of MECAT, whereas PAFI is computed on the
full benchmark. To make the two sources directly comparable, we select the
intersection between the human-rated subset and the PAFI-scored set, and
compare them at the sample level. 
For each category and each system pair, we compute paired Cohen's $d_z$ for both
human OVL and PAFI. The sign follows the order of the pair label: for a pair
A vs.\ B, we compute the signed difference as A minus B, so a positive effect
size favors A and a negative one favors B.
For example, a positive effect size for GT-EP~(GT vs.\ Expert-Pipeline) on the
OVL metric in the SM0 category indicates that human raters preferred the
ground-truth system over the Expert-Pipeline.

\Cref{fig:effect_size_agreement} visualizes the distribution of paired effect
sizes derived from human OVL and PAFI across categories and system pairs. We
evaluate consistency from two perspectives: whether the two effect sizes have
the same sign, and how strongly they correlate across all category and system pairs.

The results show that PAFI captures human preference reasonably well at the
level of relative system comparison. Across the 21 category-level pairwise
comparisons, the effect-size direction agrees in 17 cases, corresponding to an
agreement ratio of 81.0\%. In addition, the effect sizes derived from human OVL
and PAFI are positively correlated, with Pearson $r=0.8224$ and $p<0.001$.
Notably, the three overall comparisons all lie close to the ideal-fit diagonal,
further indicating that PAFI is reliable for overall system-level ranking.

Most disagreements occur in comparisons where the effect size is already close
to zero, suggesting borderline cases rather than systematic contradictions.
Moreover, 3 disagreements out of 4 occur in the Dasheng AudioGen vs.\
Ground Truth comparisons, which suggests that Dasheng AudioGen is already close
to ground truth in these categories and therefore harder to distinguish
consistently. 

Overall, these results support the use of PAFI as a useful
automatic metric for system ranking, while also indicating that human
evaluation remains necessary for resolving fine-grained differences in more
challenging cases.

\subsection{Physical Acoustic Fidelity Index (PAFI) Prompt}
\label{ssec:appendix_pafi_prompt}

\begin{tcolorbox}[
    enhanced,
    breakable,
    colback=gray!5,
    colframe=gray!70!black,
    fonttitle=\bfseries\large,
    boxrule=1pt,
    arc=4pt,
    outer arc=4pt,
    left=10pt, right=10pt, top=10pt, bottom=10pt,
    drop shadow=black!10
]

\textbf{\large Role}\\
You are a Senior Acoustic Engineer and an expert in Computational Auditory Scene Analysis (CASA). You possess profound knowledge of physical acoustics, psychoacoustics, and spatial audio coding. You are highly sensitive to physical acoustic anomalies, such as unnatural reverberation, phase cancellations, or spectral gaps, and can accurately judge the physical authenticity of an audio signal.

\vspace{1em}
\textbf{\large Task}\\
Perform a deep physical-modeling analysis of the provided audio, or of its audio description. Your objective is to determine whether the audio follows the acoustic laws of the real physical world, and whether it should be classified as an authentic capture, spliced tracks, or artifact-heavy AI audio.

\vspace{1em}
\textbf{\large Evaluation Dimensions}
\begin{enumerate}
    \item \textbf{Spatial Consistency}: Determine whether all sound sources are situated within a unified acoustic soundfield. Consider whether early reflections, distance cues, HRTF patterns, and high-frequency attenuation match the implied space.
    \item \textbf{Reverberation Coherence}: Determine whether the primary subject and background sounds share a consistent reverb profile. Strong mismatch in dryness or RT60 across sources should be treated as evidence of splicing or artificial composition.
    \item \textbf{Dynamic Layering \& Masking}: Determine whether simultaneous sources interact naturally. Consider masking effects, depth ordering, and whether overlaps sound physically plausible rather than phase-conflicting.
    \item \textbf{Environmental Immersion}: Determine whether the audio contains a credible noise floor and a coherent sense of place, including room tone, low-frequency disturbances, reflections, and diffusion.
    \item \textbf{Physical \& Kinematic Logic}: Determine whether source motion obeys acoustic physics, including Doppler shift and inverse-square SPL changes.
\end{enumerate}

\vspace{1em}
\textbf{\large Scoring Rubric (0--5)}
\begin{itemize}
    \item \textbf{0}: Severe violation. Obvious phase cancellations, fractured spectra, mismatched reverberation, or clearly broken splicing / AI artifacts.
    \item \textbf{1}: Barely acceptable. The scene is recognizable, but the acoustics are highly unstable or disconnected.
    \item \textbf{2}: Flawed. Basic acoustic logic is present, but key physical cues are missing or artifacted.
    \item \textbf{3}: Adequate. The soundfield is mostly coherent, with only mild artificial traces.
    \item \textbf{4}: Highly simulated. Physical consistency is strong and difficult to distinguish from professional production.
    \item \textbf{5}: Authentic perfection. The audio is fully coherent and indistinguishable from a high-fidelity real-world recording.
\end{itemize}

\vspace{1em}
\textbf{\large Output Format}
\begin{lstlisting}
{
  "score": <int>,
  "reason": "<string>"
}
\end{lstlisting}

\vspace{0.5em}
\textbf{\large Now, please process the following user input:}\\
\texttt{\{\{user\_input\_audio\}\}}
\end{tcolorbox}


\section{Prompt of Agentic Prompt Refiner}
\label{sec:appendix_agentic_prompt_refiner}

\begin{tcolorbox}[
    enhanced,
    breakable,
    colback=gray!5,           
    colframe=gray!70!black,   
    fonttitle=\bfseries\large,
    boxrule=1pt,
    arc=4pt,                  
    outer arc=4pt,
    left=10pt, right=10pt, top=10pt, bottom=10pt,
    drop shadow=black!10      
]

\textbf{\large Role}\\
You are an expert Audio Scene Architect and Foley Designer. Your task is to deeply understand the user's natural language audio description, and translate it into a highly structured JSON format for a \textbf{10-second audio clip}, based on real-world acoustic logic and scene realism.

\vspace{1em}
\textbf{\large Task}\\
Process the user's natural language input (which may be in any language), extract the audio elements, and output a JSON object containing the following specific keys.

\vspace{1em}
\textbf{\large JSON Schema Definition}
\begin{itemize}
    \item \texttt{Caption}: (String, Required) The overall, comprehensive description of the 10-second audio scene.
    \item \texttt{Speech}: (String, Optional) Speaker identity (e.g., middle-aged man, energetic girl) and speaking style (e.g., deep voice, anxious, echoing).
    \item \texttt{ASR}: (String, Optional) The actual transcript / spoken dialogue. \textbf{Must ONLY contain the spoken words. NO speaker labels (e.g., do not use ``Man A:'', ``Speaker 1 says:'').}
    \item \texttt{SFX}: (String, Optional) Specific sound effects present in the audio (e.g., footsteps, doorbell, dog barking).
    \item \texttt{Music}: (String, Optional) Description of background music (e.g., soft jazz, tense orchestral).
    \item \texttt{ENV}: (String, Optional) Environmental or ambient background noise (e.g., city bustle, forest wind and crickets).
\end{itemize}

\vspace{1em}
\textbf{\large Crucial Generation Rules}
\begin{enumerate}
    \item \textbf{Absolute English-Only Output}: ALL fields in the output JSON MUST be generated in English, regardless of the language of the user's input. The \texttt{ASR} field must also be entirely in English. Do NOT output any other languages.
    \item \textbf{10-Second Constraint}: The entire audio scene is exactly 10 seconds long. If generating \texttt{ASR}, keep the dialogue concise and realistic for a 10-second window (usually 1-2 short sentences).
    \item \textbf{Strict ASR Formatting}: The \texttt{ASR} field must ONLY contain the raw spoken text. Never include names, character tags, or action descriptions within the ASR string. (Correct: "Watch out for that car!" $|$ Incorrect: "Man: Watch out for that car!"). If there are multiple speakers, just combine their dialogue naturally without labels.
    \item \textbf{Scene Enrichment}: Act as a sound designer and logically enrich the scene. If the user says "at a train station," automatically add train horn (SFX) and crowd murmurs (ENV).
    \item \textbf{Logical Nulls}: Except for \texttt{Caption} (mandatory), if a field is not mentioned by the user AND makes no logical sense in the scene, set its value to \texttt{null}.
    \item \textbf{Strict Output}: Output ONLY valid JSON. Do not include markdown blocks like \verb|```json| or any explanatory text.
\end{enumerate}

\vspace{1em}
\textbf{\large Examples}

\vspace{0.5em}
\noindent \textbf{User Input}: "A man complaining about the weather on a rainy street."\\
\textbf{Output}:
\begin{lstlisting}
{
  "Caption": "A man complaining about the rainy weather on a wet city street.",
  "Speech": "A frustrated middle-aged male voice speaking loudly.",
  "ASR": "Damn it! I can't believe I forgot my umbrella again!",
  "SFX": "Heavy raindrops hitting the pavement, a car splashing water.",
  "Music": null,
  "ENV": "Urban street background noise, distant traffic."
}
\end{lstlisting}

\vspace{0.5em}
\noindent \textbf{User Input}: "两个女孩在游乐场里兴奋地指着过山车"\\
\textbf{Output}:
\begin{lstlisting}
{
  "Caption": "Two young girls expressing excitement over a roller coaster at a busy amusement park.",
  "Speech": "Two young female voices, cheerful and energetic, speaking over each other slightly.",
  "ASR": "Look at that one! It goes completely upside down! Let's go line up right now!",
  "SFX": "The mechanical clanking of a roller coaster climbing, followed by a loud whoosh.",
  "Music": "Upbeat, faint carnival music playing from nearby speakers.",
  "ENV": "Crowd chatter, distant screams of thrill, general theme park ambiance."
}
\end{lstlisting}

\vspace{0.5em}
\noindent \textbf{User Input}: "A calm ambient electronic music track for studying."\\
\textbf{Output}:
\begin{lstlisting}
{
  "Caption": "A relaxing and continuous ambient electronic music track designed for deep focus.",
  "Speech": null,
  "ASR": null,
  "SFX": null,
  "Music": "Slow, pulsing ambient synthesizer pads with a very soft, steady rhythmic beat.",
  "ENV": null
}
\end{lstlisting}

\vspace{1em}
\textbf{\large Now, please process the following user input:}\\
\texttt{\{\{user\_input\}\}}

\end{tcolorbox}

\clearpage

\section{Broader Impacts}
\label{sec:appendix_broader_impacts}
Dasheng AudioGen is designed for unified audio scene generation. It can generate complex audio scenes from text descriptions, including speech, music, sound effects, and environmental acoustics.

This technology has broad potential for positive social impact. First, it can lower the barrier to audio content creation, enabling creators in education, games, film and television, podcasts, audiobooks, and virtual reality to produce high-quality audio materials more efficiently. Compared with traditional workflows that require speech, music, and sound effects to be recorded or generated separately and then mixed in post-production, unified audio scene generation can substantially reduce production costs and provide small teams and individual creators with richer sound design capabilities. Second, this technology may also benefit accessibility applications. For example, it could be used to automatically generate immersive audio descriptions for visual content, enrich educational materials with environmental sounds and contextualized speech, or help construct more realistic training and evaluation data for auditory research, acoustic simulation, and multimodal learning systems. In addition, structured multi-view captions allow users to control speech content, music, sound effects, and environmental acoustics more explicitly, which may improve the interpretability and controllability of audio generation systems.

At the same time, unified audio generation may introduce potential societal risks. Since the model can generate realistic audio containing intelligible speech and complex background environments, misuse could enable misleading audio, synthetic media, fraud, impersonation, fabricated event recordings, or unauthorized audio content. Compared with systems that generate only a single type of audio, mixed audio scene generation may further increase the perceived realism of fabricated content and make detection more difficult. In addition, the language, geographic, cultural, and acoustic-scene distributions in the training data may be imbalanced, which could lead to inconsistent generation quality across languages, accents, cultural contexts, or sound types, and may amplify existing data biases. The generation of music and sound effects may also raise concerns related to copyright, attribution, and data provenance. Therefore, real-world deployment should carefully address training data authorization, ownership of generated content, and the boundaries of downstream use.

To mitigate these risks, we believe that such models should be deployed together with appropriate safety mechanisms and usage policies. It is important to note that the current version of Dasheng AudioGen only supports coarse-grained speaker-style control through textual descriptions. It does not support voice cloning or explicit speaker-identity conditioning. Therefore, the model itself cannot directly reproduce the voice of a specific real person. However, as unified audio generation technology continues to develop, future extensions that incorporate speaker embeddings, reference-audio conditioning, or other identity-control modules may introduce risks related to impersonation, unauthorized voice reproduction, and deceptive synthetic audio. We therefore recommend that any future extension or practical deployment avoid enabling the reproduction or impersonation of real individuals without explicit authorization, and that it incorporate synthetic-audio labeling, watermarking, and abuse-detection mechanisms. Public-facing applications should also include content moderation, misuse detection, and restrictions on high-risk use cases, such as fraud, political deception, fabricated evidence, or requests to impersonate specific individuals. For the research community, we further recommend that evaluations of unified audio generation models consider not only audio quality and text relevance, but also speech intelligibility, scene realism, cross-lingual fairness, copyright risks, and potential misuse.

This work is intended primarily for academic research. Its goal is to investigate modeling approaches, representation choices, and evaluation protocols for unified audio scene generation. We do not encourage or support the use of this technology for deception, impersonation, evidence fabrication, privacy violations, circumvention of consent, copyright infringement, or any other use that may cause harm to individuals, groups, or society. Any practical application based on this work should comply with applicable laws and regulations, data authorization requirements, and platform safety policies, and should clearly disclose the synthetic nature of generated content when it is disseminated to the public.

  
\end{document}